\documentclass[12pt,twocolumn]{emulateapj_Astroph}

\tightenlines

\usepackage{ulem} % AV
\usepackage{color} % AV
\usepackage{epsfig,amsmath}
\usepackage{times}

\newcount\listnorom
\newcommand\listromanDE{\global\advance \listnorom by 1
{\lowercase\expandafter{(\romannumeral\listnorom)}\ }}
\newcommand\newlistroman{\listnorom=0}

\def\lsim{\raise0.3ex
  \hbox{$<$\kern-0.75em\raise-1.1ex\hbox{$\sim$}}\,}
\def\gsim{\raise0.3ex
  \hbox{$>$\kern-0.75em\raise-1.1ex\hbox{$\sim$}}\,}
\newcommand{\CasA}{Cassiopeia A}
\newcommand{\RH}{Rankine-Hugoniot}
\newcommand{\Fermi}{{\it Fermi-LAT}}
\newcommand{\Suz}{{\it Suzaku}}
\newcommand{\HESS}{{\it HESS}}
\newcommand{\CC}{core-collapse}

\newcommand{\EnCR}{\epsilon_\mathrm{CR}}
\newcommand{\EnEsc}{\epsilon_\mathrm{esc}}
\newcommand{\Mswept}{M_\mathrm{swept}}
\newcommand{\CRnei}{CR-hydro-NEI}
\newcommand{\MCloud}{molecular cloud}
\newcommand{\ejPL}{n}
\newcommand{\SunMyr}{$\Msun$\,yr$^{-1}$}

\newcommand{\FS}{forward shock}

\newcommand{\rgIndex}{{\alpha_\mathrm{rg}}}
\newcommand{\nIndex}{\beta_n}

\newcommand{\Mshell}{M_\mathrm{shell}}
\newcommand{\rg}{r_g}
\newcommand{\BCSMz}{B_\mathrm{CSM,0}}
\newcommand{\DCSM}{D_\mathrm{CSM}}
\newcommand{\DCSMz}{D_\mathrm{CSM,0}}
\newcommand{\LCSM}{\lambda_\mathrm{CSM}}
\newcommand{\Lz}{\lambda_\mathrm{CSM,0}}

\newcommand{\Norm}{f_\mathrm{norm}} %%{N_\mathrm{DSA}}
\newcommand{\NCSM}{n_\mathrm{CSM}}

\newcommand{\Nuni}{n_\mathrm{uni}}
\newcommand{\Nshell}{n_\mathrm{shell}}
\newcommand{\Bshell}{B_\mathrm{shell}}

\newcommand{\Rshell}{R_\mathrm{shell}}

\newcommand{\dMdt}{dM/dt}
\newcommand{\Vwind}{V_\mathrm{wind}}
\newcommand{\Twind}{T_\mathrm{wind}}
\newcommand{\SigWind}{\sigma_\mathrm{wind}}
\newcommand{\cutoff}{\alpha_\mathrm{cut}}
\newcommand{\Ncut}{N_\mathrm{cut}}

\newcommand{\pcc}{cm$^{-3}$}
\newcommand\Msun{\mathrm{M}_{\odot}}

\newcommand{\tempEq}{f_\mathrm{eq}}

\newcommand{\Bamp}{B_\mathrm{amp}}
\newcommand{\EffDSA}{{\cal E_\mathrm{DSA}}}

\newcommand{\Kep}{K_\mathrm{ep}}

\newcommand\tSNR{t_\mathrm{SNR}}
\newcommand\dSNR{d_\mathrm{SNR}}
\newcommand\EnSN{E_\mathrm{SN}}
\newcommand\Mej{M_\mathrm{ej}}
\newcommand{\RFS}{R_\mathrm{FS}}

\newcommand{\fsk}{f_\mathrm{sk}}
\newcommand{\VelFS}{V_\mathrm{FS}}

\newcommand{\SA}{semi-analytic}
\newcommand{\NT}{non-thermal}
\newcommand{\NEI}{non-equilibrium ionization}
\newcommand{\CD}{contact discontinuity}
\newcommand{\xx}[1]{\!\times\!10^{#1}}

\newcommand{\DSA}{diffusive shock acceleration}
\newcommand{\CSM}{circumstellar medium}

\newcommand{\rgz}{r_{g,0}}

\newcommand{\kmps}{km s$^{-1}$}
\newcommand{\NL}{nonlinear}
\newcommand{\gamray}{$\gamma$-ray}
\newcommand{\gamrays}{$\gamma$-rays}

\newcommand{\SNRJ}{SNR RX J1713.7-3946}

\newcommand{\SNRJmm}{SNR J1713}
\newcommand{\SC}{self-consistent}
\newcommand{\SCly}{self-consistently}
\newcommand{\Rtot}{r_\mathrm{tot}}

\newcommand{\muG}{$\mu$G}

\newcommand{\be}{\begin{eqnarray}}
\newcommand{\ee}{\end{eqnarray}}

\newcommand{\DSAinj}{\chi_\mathrm{inj}}

\newcommand{\rel}{relativistic}

\newcommand{\mc}{Monte Carlo}

\newcommand{\syn}{synchrotron}
\newcommand{\synch}{synchrotron}
\newcommand{\pion}{pion-decay}
\newcommand{\IC}{inverse-Compton}
\newcommand{\brem}{bremsstrahlung} % DCE
\newcommand{\brems}{bremsstrahlung} % DCE
\newcommand{\red}{\textcolor{red}}

\newcount\Inum
\newcount\IInum
\newcount\IIInum
\newcount\IVnum
\def\I{\global\multiply\IInum by 0 \global\multiply\IIInum by 0
            \global\multiply\IVnum by 0 \global\advance \Inum by 1
            {\the\Inum. }}
\def\II{\global\multiply\IIInum by 0\global\multiply\IVnum by 0
       \global\advance \IInum by 1 {\the\Inum.\the\IInum. }}
\def\III{\global\multiply\IVnum by 0\global\advance \IIInum by 1
            {\the\Inum.\the\IInum.\the\IIInum. }}
\def\IV{\global\advance \IVnum by 1
            {\the\IVnum. }}
\begin{document}

\title{Core-collapse model of broadband emission from \SNRJ\ with
thermal X-rays and Gamma-rays from escaping cosmic rays}

\author{Donald C. Ellison,\altaffilmark{1}
Patrick Slane,\altaffilmark{2} Daniel
  J. Patnaude,\altaffilmark{2} and Andrei M. Bykov
\altaffilmark{3}}

\altaffiltext{1}{Physics Department, North Carolina State
University, Box 8202, Raleigh, NC 27695, U.S.A.;
don\_ellison@ncsu.edu}

\altaffiltext{2}{Smithsonian Astrophysical Observatory, MS-3, 60
Garden
  Street, Cambridge, MA 02138, USA}

\altaffiltext{3}{Ioffe Institute for Physics and Technology, 194021
St. Petersburg, Russia; byk@astro.ioffe.ru}

\begin{abstract}
We present a spherically symmetric, core-collapse model of SNR RX
J1713.7-3946 that includes a hydrodynamic simulation of the remnant
evolution coupled to the efficient production of cosmic rays (CRs) by
\NL\ \DSA\ (DSA). High-energy CRs that escape from the
forward shock (FS) are propagated in surrounding dense material that
simulates either a swept-up,
pre-supernova shell or a nearby molecular cloud. The continuum
emission from trapped and escaping CRs, along with the thermal X-ray
emission from the shocked heated ISM behind the FS, integrated over
the remnant, is compared against broadband observations. Our results
show conclusively
that, overall, the GeV-TeV emission is dominated by \IC\ from CR
electrons if the supernova is isolated regardless of its type, i.e.,
not interacting with a $\gg\! 100\,\Msun$ shell or cloud.
If the SNR is interacting with a much larger mass
$\gsim 10^4\,\Msun$, \pion\ from the escaping CRs may dominate the TeV
emission, although a precise
fit  at high energy will depend on the still uncertain details of how
the highest energy CRs are accelerated by, and escape from, the FS.
Based on morphological  and other constraints, we consider the
$10^4\,\Msun$ \pion\ scenario highly unlikely for \SNRJ\ regardless of
the details of CR escape. Importantly, even though CR electrons
dominate the
GeV-TeV emission, the efficient production of CR  {\it\large ions}
 is an essential part of our leptonic model.
\end{abstract}

\keywords{ acceleration of particles, shock waves, ISM: cosmic rays,
           ISM: supernova remnants, magnetic fields, turbulence}

\section{Introduction}
The direct observation of radio, \NT\ X-ray, and GeV-TeV emission from
several young supernova remnants (SNRs) provides unambiguous proof
that these objects produce \rel\ particles. The morphology of this
emission, which is often seen in thin, rim-like structures associated
with the supernova (SN) blast wave,  provides convincing evidence that
\DSA\ (DSA) is the mechanism producing these particles. This and other
evidence, such as the direct measure of efficiency at the Earth bow
shock \citep[e.g.,][]{EMP90}, and the likely existence of amplified
magnetic fields in some SNRs,
suggests that the efficiency for producing superthermal ions, i.e.,
cosmic rays (CRs), is high in DSA.
A critical issue for DSA, and for the origin of CRs, however, concerns
the
radiation mechanism responsible for the GeV-TeV emission. Is it
dominated by \pion\ emission from ions or  \IC\ emission from leptons?

Given \rel\ electrons and ions, three mechanisms -- \NT\ \brem,
\IC\ (IC) scattering, and \pion\ -- can produce GeV-TeV photons.
Sorting out this mix in a particular remnant not only provides
information on how CRs are produced, it constrains the basic physics
of DSA, in particular the electron to proton ratio,
$\Kep$, of accelerated
particles. While most modelers of young remnants claim that  shock
accelerated protons are mainly responsible for the GeV-TeV emission
\citep[e.g.,][]{MAB2009,BV2010}, some researchers
\citep[e.g.,][]{PMS2006,KW2008,ZirA2010},  have suggested that
IC from \rel\ electrons dominates the GeV-TeV emission from
\SNRJ\ (henceforth \SNRJmm).
Recently we have definitively shown,  with a model
that includes the production of thermal X-ray lines \SCly\ with the
broadband continuum emission \citep[i.e.,][]{EPSR2010}, that IC
emission from \rel\ electrons clearly dominates \pion\ if \SNRJmm\ is
in a homogeneous environment
typical of a Type Ia SN.

The assumption of a homogeneous environment, although typically
assumed in \SC\ models of \SNRJmm, is an important caveat because
\SNRJmm\  is more
likely a core-collapse SN that may have exploded in a complex
environment
\citep[e.g.,][]{Fukui2003,Moriguchi2005,Inoue2009,Sano2010}.
Because of this, we have generalized our model in two important ways.
First, we can now model a spherically symmetric, non-homogeneous
environment. We consider a SN that explodes in a pre-SN wind
surrounded by a dense, swept-up shell of wind material. We further
allow for external dense material that might be associated with
molecular clouds out of which massive stars form. Second, in addition
to \gamrays\ produced by CRs trapped within the \FS\ (FS), we now
include \gamray\ production from the CR protons  that escape from the
FS and diffuse rapidly into the dense shell and/or the molecular cloud
material beyond the forward shock.

These generalizations allow a fairly realistic model of a
core-collapse \SNRJmm\ interacting with dense material, at the expense
of adding model parameters. However, despite the added parameters and
the flexibilty in fitting they allow, we find it is impossible to
produce a good fit to the broadband data, including the X-ray emission
lines, with  \pion\ dominating \IC\ at GeV-TeV energies unless the SNR
is interacting with a dense mass concentration ($\gsim 10^4\,\Msun$)
such as a molecular cloud \citep[e.g.,][]{GA2007,GAC2009}. An isolated
SNR,
regardless of its type,
even allowing for escaping CRs interacting with a dense shell of
material with $\Mshell\, \lsim 100\,\Msun$ from a pre-SN wind, is
excluded.

The conclusion of  \citet{EPSR2010}, that \IC\ from electrons
dominated the GeV-TeV emission, hinged on the assumption that both \NT
\ \pion\ and thermal X-ray line emission scale in approximately
the same way with the ambient density. This is expected to be the case
if the shock accelerated particles are drawn from the target material,
i.e., if the material swept-up by the FS provides seed particles for
DSA and simultaneously provides a dense target for proton-proton
collisions. If the environment is homogeneous and has a uniform
density everywhere outside of the FS, as assumed in \citet{EPSR2010},
photons produced outside of the FS, where the density is considerably
less than behind the shock, can be ignored. In this case, the
parameters chosen for the injection of electrons and protons into DSA,
along with the ambient density and magnetic field, largely
determine the IC/\pion\ ratio and the thermal X-ray emission/\pion\
ratio. As shown by
\citet{EPSR2010}, reasonable parameters for \SNRJmm\ exclude \pion\ as
the dominant radiation mechanism for the GeV-TeV emission in this
scenario.

There are three main populations of
shock accelerated CRs that are important for producing \gamrays:
relativistic electrons producing \gamrays\ through \IC\ and \NT\
\brems; CR ions that remain trapped within the \FS; and CR ions that
are accelerated by the \FS\ but escape upstream.\footnote{A fourth
particle population that we don't consider here are secondary
electron-positron pairs produced by proton-proton interactions
\citep[see, for example,][]{GAC2009}. These leptons will produce IC
emission and may be important depending on the external mass
concentration.}
These three populations are produced simultaneously by DSA in the
blast wave and, while they have different properties, their radiation
signatures, particularly IC and \pion, are not easily distinguished at
GeV-TeV energies with current observations. The recent \Fermi\
observations of \SNRJmm,  however, now
clearly support a leptonic model for the GeV-TeV emission
\citep[][]{AbdoJ1713_2011}.

An important distinction between trapped and escaping CR ions is
that, in a non-homogeneous environment, where regions outside of the
FS can have an elevated density, the escaping CR ions can diffuse
into these dense regions and produce \gamrays\ without producing
any corresponding thermal X-ray or IC emission.\footnote{For the
purposes of this paper, we assume that only CR ions escape upstream
from the shock. Relativistic electrons will suffer radiation
losses and are not as likely to be an important contributor to the
IC
emission beyond the FS. Including escaping CR electrons would increase
the
IC/pion-decay ratio from the values we calculate and strengthen our
basic conclusions.}
The production of these external \gamrays\ depends mainly on the
amount of external mass but the diffusion properties of the CRs in the
\CSM\ (CSM) also come into play. In \citet{EB2011} we presented a
simple
\mc\ model
for escaping CR diffusion and investigated how the \gamray\ flux
depends on the CR diffusion without application to a particular SNR.
For reasonable values of the diffusion, it is clear that, if the
escaping CRs have enough external mass to interact with, the \gamray\
flux from \pion\ can overwhelm that from IC
\citep[e.g.,][]{GA2007,LKE2008}. Other observational
considerations, such as the existence of nearby molecular clouds and
the spatial coincidence of radio, X-ray \syn, and \gamray\ emission
must be used to discriminate between a hadronic or leptonic origin in
this case.

Here, we model the broadband emission from \SNRJmm\ for
four  non-homogenous environments.
The first is for an isolated core-collapse SNR expanding in a slow
pre-SN wind with an external, dense shell of swept-up wind material.
For an isolated SNR, the mass of the swept-up wind must be less than
the mass of the main sequence star and we use $\Mshell = 100\,\Msun$
as an extreme case.
In the second example, we increase the external mass to
$\Mshell = 10^4\,\Msun$ to minmic a SNR which is near a mass
concentration but has not yet impacted it.
In the third example, we allow the SNR to impact a dense shell of
$10^4\,\Msun$ and follow the evolution of the remnant and the photon
production as the FS moves into the shell.
Finally, we consider the case of a fast pre-SN wind.

We find, largely independent  of diffusion parameters, that escaping
CRs interacting with an external mass shell from a pre-SN wind with
mass
$\lsim 100\,\Msun$ will not produce enough \pion\ emission to
compete with the IC emission from trapped CR electrons. If larger
external masses $\gsim 10^4\,\Msun$ are allowed, as would be the case
for a SNR near or
interacting with a \MCloud, the \pion\ can dominate the IC at TeV
energies. Now, however, other considerations come into play such as
the shape of the escaping CR distribution and the details of how CRs
can stream into a molecular cloud and produce self-generated
turbulence. We expect the escaping CR distribution to be far too
narrow to allow a good fit to the GeV-TeV observations without
a substantial contribution from IC photons.
Nevertheless, the shape is uncertain since escaping CRs
 will be
highly anisotropic and the plasma physics of these distributions is an
active area of research.\footnote{We note that in addition to the
direct escape of CRs upstream from the shock that we consider, an
additional factor comes into play in an expanding SNR. As the remnant
expands, the precursor region
beyond the forward shock that is filled with CRs expands producing a
``dilution" of CR energy density.  This effect has been studied in
detail by Berezhko and co-workers \citep[e.g.,][]{BEK1996a,BEK1996b}
\citep[see also][]{Drury2010}. In a real shock, the dilution effect is
coupled to escape since the lowering of the CR energy density results
in less efficient generation of magnetic turbulence and this will
change the escape of CRs. Both the flow of energy out of the shock by
escape and the dilution of the CR energy density influence the \RH\
conservation relations in similar ways. Both act as energy sinks and
both result in an increase in the shock compression and other \NL\
effects.}

The broadband fit we obtain for our IC \CC\ model
is, in fact, significantly better than that obtained in
\citet{EPSR2010}, particularly for the highest energy \HESS\ points.
The
reason, as we show below, stems from the fact that the
pre-SN wind magnetic field, in which the SN explodes, can be
considerably
lower than 3\,\muG, allowing electrons to be accelerated to higher
energies before radiation losses dominate.
%Even with substantial MFA, the shocked field is
%lower than in the homogeneous case and
%electrons are accelerated to higher energies without suffering
%radiation losses, allowing a better fit to the \HESS\ points.

It is important to emphasize that our model, and all \SC\ shock
acceleration models of \SNRJmm\ we are aware of, accelerate CR
{\it\large ions} efficiently. The models we show below
 place 25 to 50\%
of the FS ram
kinetic energy flux into \rel\ ions at any
instant. Only 0.25\% or less of the
instantaneous ram kinetic energy
flux goes into \rel\ electrons. Leptons dominate the observed emission
simply because leptons radiate far more efficiently than ions, not
because ions are missing. Furthermore, our best-fit parameters for
this remnant result in maximum
proton energies of $\sim 10^{14}$\,eV. Iron nuclei would be
accelerated to $\sim 26\xx{14}$\,eV, well into the CR ``knee." We make
no claim that the spectral shape produced in this particular SNR
matches the CR spectrum observed at Earth. As has been known for
decades \citep[e.g.,][]{Ginzburg1964}, CRs observed at Earth come from
a mix of many SNRs, with some 
%possible small 
admixture of other sources \citep[see][]{MDE97}, after a long
and tortuous propagation in the ISM. The spectrum produced by any one
source will be modifed substantially by propagation
\citep[see figure 1 in][]{EE85}, and different elements, e.g.,
hydrogen and helium, may propagate differently
\citep[e.g.,][]{BA2011a}.
%and uncertainties in propagation should be considered before calling
%into question SNRs as the primary source of galactic CRs care must be
%taken if this spectrum is  compared directly to the spectrum observed
%at Earth.
Despite some recent claims \citep[][]{AdrianiEtal2011}, we believe our
detailed fit to \SNRJmm\ is fully consistent with SNRs being the
primary source of galactic CR ions at least to energies into the CR
knee.

\section{Model}
\label{sec:model}
The cosmic-ray, hydrodynamic, \NEI\ (CR-hydro-NEI) model of an
evolving SNR we use has been described in detail in a number of
previous papers. It  is essentially the same as that described in
\citet{EPSR2010} and references therein except that we  have added the
escape of high-energy CRs from the shock precursor and the propagation
and \gamray\ production of these CRs as they diffuse in the \CSM. Our
method describing the escaping CRs is given in  \citet{EB2011}.

\newlistroman

The important  features of the \CRnei\ model are:
\listromanDE the hydrodynamics and evolution of the SNR are coupled to
the efficient production of CRs via \NL\ DSA;
\listromanDE the NL CR acceleration is calculated using the \SA\
technique given in \citet{BGV2005};
\listromanDE the CRs that remained trapped within the FS have a normalization and maximum energy cutoff consistent with the normalization and shape of the high-energy CRs that
escape upstream;
\listromanDE magnetic field amplification (MFA),
presummably produced by
\NL\ DSA at the FS,  is simply parameterized;
\listromanDE as described in detail in \citet{EPSBG2007}, the NL acceleration model determines the heating
of the shocked plasma;
%\sout{the NL acceleration model \SCly\ determines the heating
%of the shocked plasma;}
%
\listromanDE the electron temperature and \NEI\ state of the shocked
plasma are calculated and  followed
\SCly\ with the SNR dynamics in the interaction region between the \CD\ and the FS as the
SNR evolves;
\listromanDE within the SNR, thermal X-ray emission lines and
continuum emission from \syn, \IC, \brems, and \pion\ from trapped CR
electrons and protons are calculated  taking into account the SNR
evolution and adiabatic and radiation losses;\footnote{As in previous
applications of our \CRnei\ model, we only calculate X-ray emission
from the shocked ISM. If thermal emission from the shocked ejecta
material was included, the X-ray line emission we show would be
greater and our basic conclusion concerning the dominance of IC
emission over \pion\ at GeV-TeV energies would be strengthened.}
and
\listromanDE the \pion\ emission from escaping CR protons and trapped CRs is absolutely normalized relative to each other and relative to the continuum and line emission in all other bands.
%the is calculated
%\SCly\ with the emission from the trapped CRs.}
%
%\sout{\pion\ emission from escaping CR protons is calculated
%\SCly\ with the emission from the trapped CRs.}

Since, as in our previous applications of the CR-Hydro-NEI model, we
assume spherical symmetry, all of our escaping CRs interact with the
external shell. In 3-D geometry, where the external mass is
concentrated in clumps, only some fraction of the escaping CRs would
encounter the clumps before diffusing away
\citep*[see, for example,][]{LKE2008}. Our \gamray\ fluxes from
escaping CRs interacting with a given external mass are therefore
upper limits.

Any realistic description of a SNR requires a large number of
parameters. Parameters are required to model the supernova explosion
and SNR,
the DSA mechanism, the CSM, and the diffusion of escaping CRs in the
CSM.
In  Tables~\ref{tab:tableSNR}, \ref{tab:tableDSA}, and
\ref{tab:tableCSM} we list these input parameters, and in
Table~\ref{tab:tableOUT},
we list some output values.

Besides the assumed age and distance to \SNRJmm,
Table~\ref{tab:tableSNR} gives values for the SN explosion energy
$\EnSN$, and ejecta mass $\Mej$, as well as the power-law index of the
initial ejecta mass distribution, $\ejPL$
\citep[see][for a full discussion of $\ejPL$]{EDB2004}.
All of our examples here assume a pre-SN wind with a speed $\Vwind$,
mass-loss rate $\dMdt$, and temperature $\Twind$, all of which are
assumed constant. The parameter $\SigWind$ shown in
Table~\ref{tab:tableSNR}, determines the wind magnetic field at a
radius $R$ from the explosion according
to
\begin{equation}\label{eq:sigma}
B_0(R) = \frac{(\SigWind \Vwind \dMdt)^{1/2}}{R}
\ ,
\end{equation}
as in \citet{CL94} and \citet{EC2005}.
The constant parameter $\SigWind$ is the ratio of magnetic field
energy density
to kinetic energy density in the wind and can be related to
properties of the star by \citep[e.g.,][]{walderea11}
\begin{equation}\label{eq:sigmaTwo}
\SigWind \propto \frac{B_{\ast}^2 R_{\ast}^2 }{(\dMdt)v_w}\times
\left(\frac{v_r}{v_w}\right)^2
\ .
\end{equation}
Here, $B_{\ast}$ denotes the surface magnetic field of the star,
$R_{\ast}$ is the stellar radius, $v_r$ is the star rotation velocity,
and $v_w=\Vwind$ is the terminal speed of the wind.
Obtaining $\SigWind \sim 0.03$ by fitting the
spectrum of \SNRJmm\ as we do, constrains the progenitor star
parameters.
As noted by \citet{walderea11}, values of $\SigWind \ll 1$ indicate
that the stellar wind dominates
the magnetic field, producing a roughly radial field far from the
star.

All of the parameters in Table~\ref{tab:tableDSA} have been described
previously \citep[see][and references therein]{EPSR2010}. Briefly,
$\EffDSA$ is the fraction of ram kinetic energy flux that is placed in
superthermal particles at any instant,\footnote{We note that we
determine the
acceleration efficiency in the \citet{BGV2005} \SA\ calculation by
setting $\EffDSA$ and then calculating the injection
fraction $\DSAinj$ rather
than the reverse, as is typically done
\citep[see][]{EPSR2010}. With a constant $\EffDSA$, $\DSAinj$ varies
over the lifetime of the SNR.}
$\Bamp$ is an amplification factor for the shocked magnetic field,
$\Kep$ is the ratio of electrons to protons at \rel\ energies, $\fsk$
is the fraction of the FS radius used to determine the maximum CR
momentum, $\cutoff$ is a factor that determines the shape of the
particle cutoff at high momentum for trapped CRs, and
$\Ncut$ (along with  $\cutoff$) determines the shape of the  escaping
CR distribution
\citep[see][for a detailed discussion of $\cutoff$ and
$\Ncut$]{EB2011}.
The parameter $\tempEq=1$ indicates that Coulomb equilibration is used
for electron heating while $\tempEq=0$ indicates that electrons
equilibrate instantly with protons downstream from the shock
\citep[see][]{EPSBG2007}, and $\Norm$ is the overall
normalization applied to the photon emission to match the
observations. In all cases, a solar
elemental composition has been assumed for the CSM.

\begin{figure}
\epsscale{0.99}
\plotone{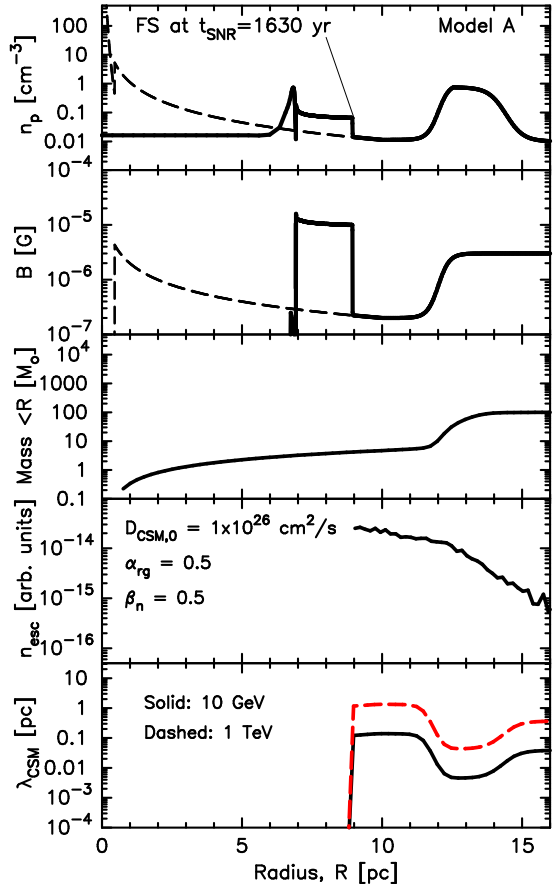}          % Fig 1
\caption{The top two panels show the proton number density and the
magnetic field as a function of distance from the center of the SNR
for Model $A$. In each of these two panels, the dashed curve is the
profile at the beginning of the simulation and the solid curve is the
profile at $\tSNR=1630$\,yr. The third panel shows the mass within $R$
at $t=0$ and the fourth panel shows the escaping CR number density.
The diffusion parameters as defined in  Eq.~\ref{eq:mfp} are listed in
the fourth panel. Escaping CRs are only followed beyond the FS and
they leave the spherically symmetric simulation freely at the outer
radius of  $\sim 16$\,pc. The sharp dropoff in $\LCSM$ within
$\sim 9$\,pc indicated in the bottom panel shows the effect of
assuming Bohm diffusion for the trapped CRs.
\label{fig:profA}}
\end{figure}

The values given in Table~\ref{tab:tableCSM} determine the
characteristics of the CSM and the diffusion of the escaping CRs.
As described in \citet{EB2011}, the diffusive mean free path for the
escaping CRs in the CSM is parameterized by
\begin{equation} \label{eq:mfp}
\LCSM = \Lz (\rg/\rgz)^\rgIndex [\NCSM(R)/n_0]^{-\nIndex}
\ .
\end{equation}
Here, $\rg=pc/(eB)$ is the gyroradius, $\NCSM(R)$ is the CSM proton
number
density,\footnote{The density $\NCSM$ is the value of $n_p$ shown in
Figs.~\ref{fig:profA}, \ref{fig:profB}, \ref{fig:Time14_18},
and \ref{fig:ShellDist} at a given $R$ outside of
the FS.}
and $\rgIndex$ and $\nIndex$ are parameters.
For scaling, we
use $n_0 = 1$\,\pcc, $\rgz = 10 \mathrm{GeV}/(e \BCSMz)$, and
$\BCSMz = 3$\,\muG.
We note that instead of the $\NCSM$ term, we could have scaled $\LCSM$
with $(B/\BCSMz)^{-\nIndex}$, or even a combination of density and
magnetic field terms. These are essentially equivalent
parameterizations unless the connection between background field,
ambient density, and wave generation by streaming CRs is specified.
The normalization of the CSM diffusion coefficient,
$\DCSMz= \Lz \, c/3$, can be estimated from CR propagation studies
\citep[see, for example,][]{PMJSZ2006,GAC2009}. For example, with
$\DCSMz=10^{27}\,$cm$^2$ s$^{-1}$, $\NCSM=0.01$\,\pcc,
$\rgIndex=0.5$, and
$\nIndex=1$, $\LCSM \sim 1$\,pc at 1 GeV, consistent with the fits of
\citet{PMJSZ2006}.
In general, the stronger the diffusion (i.e., the smaller $\LCSM$) the
greater the \gamray\ emission will be in the external material.

\begin{figure}
\epsscale{0.99}
\plotone{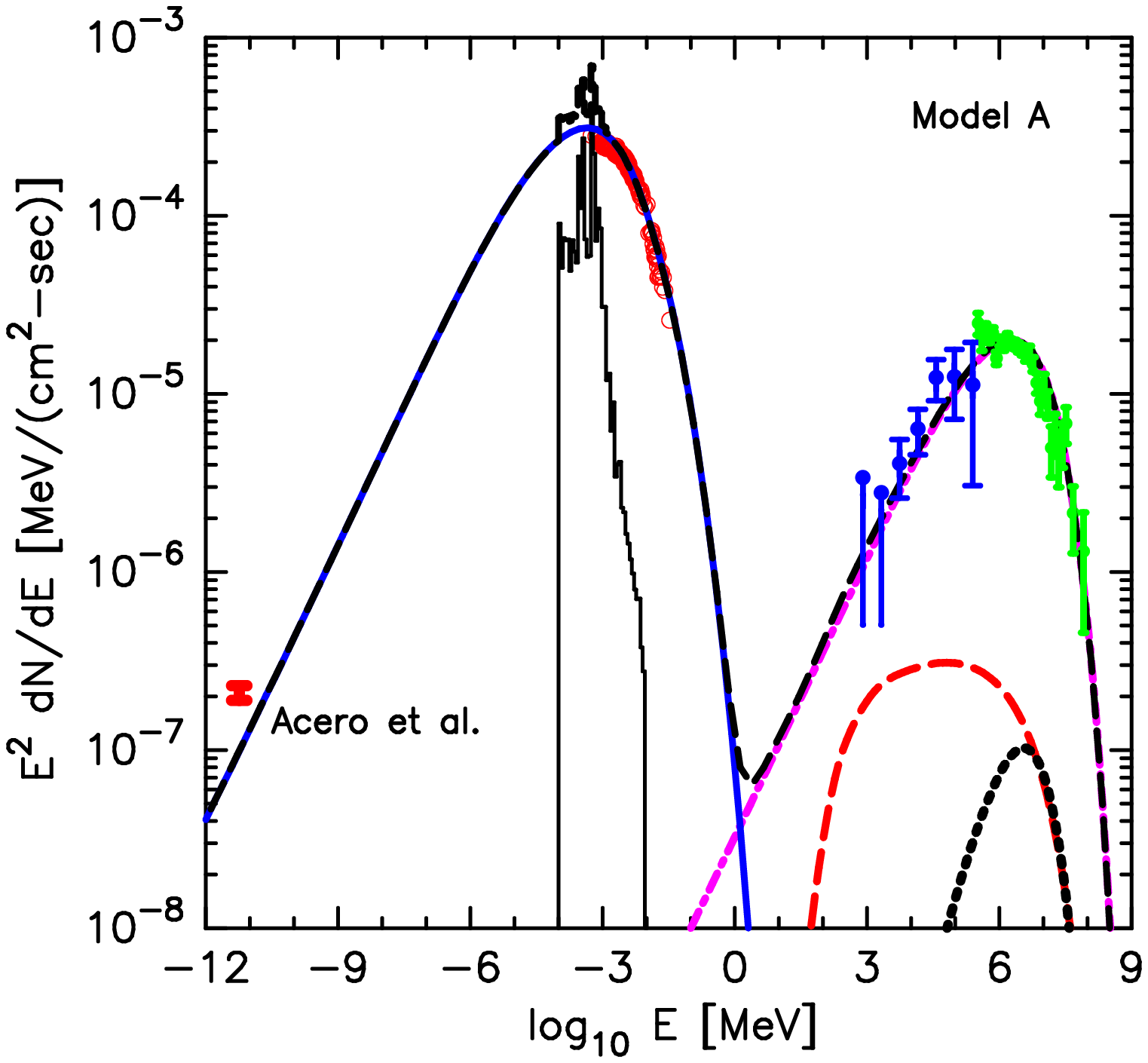}          % Fig 2
\caption{Model $A$ fit to \SNRJmm\ observations. The different
emission
processes are: \syn\ (solid blue curve), IC (dot-dashed purple curve),
\pion\ from trapped CRs (dashed red curve), \pion\ from escaping CRs (
dotted black curve), and thermal X-rays (solid black curve). The
dashed black curve is the summed emission. The data is from
\citet{Acero2009} (radio), \citet{Tanaka2008} (\Suz\ X-rays),
\citet{AbdoJ1713_2011} (\Fermi), and
\citet{Aharonian_J1713_ERR2011} (\HESS). Note that the two lowest energy
\Fermi\ points are upper limits. For all models we use a column
density of $n_H = 7.9\xx{21}$\,cm$^{-2}$.
\label{fig:BbandA}}
\end{figure}

The values  $\Nuni$ and $\Nshell$ are the proton number
densities for the
uniform CSM beyond the dense shell and for the dense shell,
respectively.
The values $\Mshell$ and $\Rshell$ are the mass of the dense shell and
its inner radius, respectively, and $\Bshell$ is the magnetic field in
the shell. As shown in Figs.~\ref{fig:profA} and \ref{fig:profB}, we
smooth the transition from the pre-SN wind to the dense shell.

We note that within the FS we assume Bohm diffusion for the CRs with a
mean free path $\lambda \sim \rg$ which is very much smaller than
$\LCSM$. This is reflected in the sharp drop in $\lambda$ within the
FS as shown in the bottom panels of Figs.~\ref{fig:profA} and
\ref{fig:profB}.

\section{Results}
\subsection{Pre-SN Wind Interaction}
For our core-collapse model A, we take the SN explosion
energy to be $\EnSN=10^{51}$\,erg, the ejecta mass $\Mej=3\,\Msun$,
and assume a slow, dense, pre-SN wind
%, such as might occur with a
%red-supergiant,
with  a mass-loss rate
$\dMdt = 10^{-5}$\,\SunMyr, and wind speed $\Vwind=20$\,\kmps.
Our model is a simplified
description that might resemble what happens after an early-type star
with a fast wind creates a large, low-density bubble before evolving
into a red-supergiant with a much slower
wind \citep[see, for example,][]{Chevalier99}.
As we show below, the critical conditions that result in a good
leptonic
fit are that the
density is relatively low and the B-field in the wind is lower than
the normal ISM
field. Other than this, none of our conclusions depend critically on
particular wind parameters.

To determine the unshocked
magnetic field as a function of radius, $R$,
in the pre-SN wind, we take $\SigWind=0.03$ in
equation~(\ref{eq:sigma}).
At the assumed age of \SNRJmm\ (i.e., $\tSNR \simeq 1630$\,yr),
the FS
has not yet reached the dense material of the swept-up wind. The
situation is shown in Fig.~\ref{fig:profA} where, in the top two
panels, the  proton number density, $n_p$, and the magnetic field,
$B$, are plotted as functions of radius, $R$, from the center of the
SNR. The dashed curves in the top two panels are the density and
magnetic field profiles at the start of the simulation. The solid
curves are these profiles at $\tSNR=1630$\,yr. Parameters have been
chosen so the SNR radius is $\sim 9$\,pc at $\tSNR=1630$\,yr,
consistent with a distance to \SNRJmm\ of $\sim 1$\,kpc and an angular
size of $\sim 60$\,arcmin.
As seen in the second panel, the pre-SN wind magnetic field just
upstream of the FS, as determined by $\SigWind$,
is $\sim 0.2$\,\muG\ at $\tSNR=1630$\,yr and this is increased by
compression and amplification to
$\sim 10$\,\muG\ immediately downstream.

At a radius beyond the FS, we have placed a dense shell with a total
mass $\sim 100\,\Msun$ and the third panel in Fig.~\ref{fig:profA}
gives the mass within $R$. A shell mass of $100\,\Msun$ is an extreme
case and any shell from swept-up, pre-SN wind material would be
considerably less. The fourth  panel in Fig.~\ref{fig:profA} shows the
number density of escaping CRs that have diffused beyond the FS and
the bottom panel shows the scattering mean free path beyond the FS,
$\LCSM$, for 10\,GeV (solid black curve) and 1\,TeV (dashed red curve)
CRs. Note that we have set $B=3$\,\muG\ in the shell and beyond.
Since we don't calculate the \synch\ emission from
electrons beyond the FS, the value of $B$ beyond the FS is
unimportant.

In Fig.~\ref{fig:BbandA} we show the radiation produced by the trapped
and escaping CRs, along with the observations of \SNRJmm\ at radio
\citep[][]{Acero2009}, X-ray \citep[Suzaku;][]{Tanaka2008},
GeV \citep[\Fermi;][]{AbdoJ1713_2011}, and TeV energies
\citep[\HESS;][]{Aharonian2007}.
As described in \citet{EPSR2010}, this is a ``best fit" result where
we
have varied parameters to obtain a good match to the broadband data.
\newlistroman
The essential features of the comparison are:
\listromanDE \IC\ emission from \rel\ electrons interacting with the
cosmic microwave background (CMB) strongly dominates the GeV-TeV
emission;
\listromanDE the quality of this broadband fit with a pre-SN wind,
including weak X-ray emission lines consistent with the \Suz\
observations, is {\it\large superior} to that obtained with a uniform
CSM
model by
\citet{EPSR2010};
\listromanDE the escaping CRs (black dotted curve), even though they
diffuse through a region with $100\,\Msun$ of material, give a small
contribution to the TeV \pion\ emission, and;
\listromanDE even though there are a lot of relatively free parameters
for this fit, the critical result that IC dominates over \pion\ at
GeV-TeV energies is robust.

We don't show a fit where parameters have been optimized for \pion\
dominance at GeV-TeV energies because the result is essentially the
same as shown in \citet{EPSR2010} for the homogeneous case.
As is clear from Fig.~\ref{fig:BbandA}, if $\Kep$ is decreased
to allow \pion\ to dominate IC, the overall density will have to be
increased substantially to produce enough flux at GeV-TeV energies.
Any increase
in density (here, for the pre-SN wind through varying $\dMdt$ and/or
$\Vwind$) will increase the X-ray line emission along with the \pion\
emission, producing a conflict with the \Suz\ observations.
In fact, as long as the contribution from the escaping CRs remains
well below the IC, all fitting parameters, such as the efficiency for
DSA, $\EffDSA$, are constrained here just as in the fits shown in
\citet{EPSR2010}.

We note that the parameter $\fsk$ determines the precursor length in our model but that we do not explicitly model the spatial properties of the precursor in our implementation of the Blasi et al. DSA model \citep[see][and references therein, for an updated version of the model that does include the precursor explicitly]{CBAV2009}. 
We do, however, account for all CRs that are accelerated and, except for the escaping CRs,  assume they are all trapped behind the FS. We, therefore, overestimate the \pion\ flux since the trapped CRs interact with the dense shocked plasma rather than the thinner precursor material. The only possible case where the precursor CRs might enhance the \pion\ emission is if the precursor is interacting with external material denser than the shocked plasma. We consider this case unlikely because it requires fine tuning. If the external material is fully outside of the precursor, it is contained in the case we discuss in Section~\ref{sec:MC} below. If the FS is also interacting with the external material, it is the case we discuss in Section~\ref{sec:MCimpact} below. 
Only if the  precursor, but not the FS,  is impacting dense material is there a possibility that the \pion\ emission will be greater than we estimate. With a precursor length determined by $\fsk = 0.1$, we consider this to be unlikely. While larger values of $\fsk$ are possible, they imply strong self-generated turbulence far upstream where the density of accelerated CRs has dropped significantly from that at the shock \citep[this is, in effect, CR ``dilution" as discussed by][]{BEK1996a,BEK1996b}. It is noteworthy that neglecting the spatial properties of the precursor is less likely to overestimate the electron contribution to the GeV-TeV emission since radiation losses will prevent the highest energy electrons from streaming far upstream.

\begin{figure}
\epsscale{0.99}
\plotone{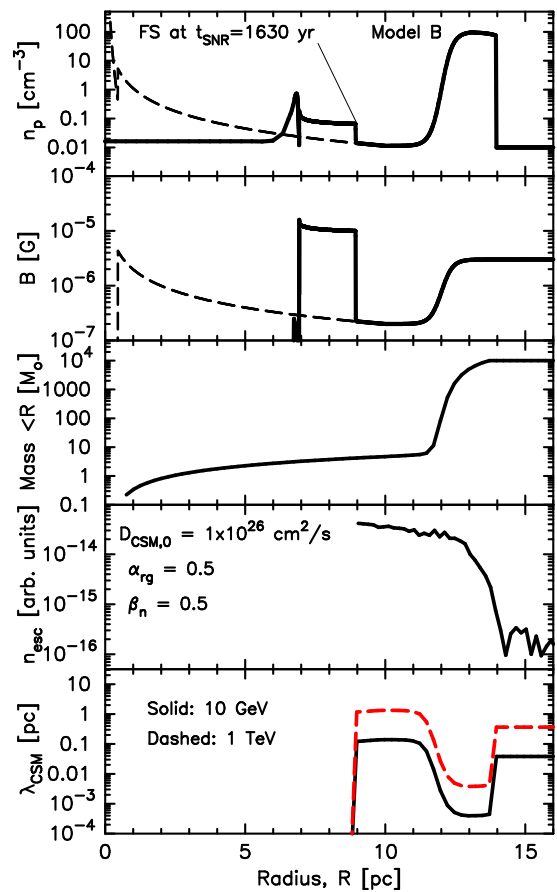}          % Fig 3
\caption{Same format as in Fig.~\ref{fig:profA} for Model $B$, where
the mass of the external shell is $\sim 10^4\,\Msun$.
\label{fig:profB}}
\end{figure}

\begin{figure}
\epsscale{0.99}
\plotone{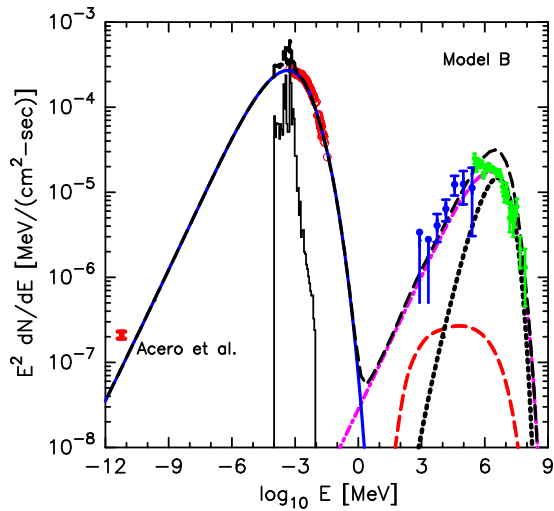}          % Fig 4
\caption{Same format as in Fig.~\ref{fig:BbandA} for Model $B$.
As in Fig.~\ref{fig:BbandA}, the black dashed curve is the total
emission and, in this case, it lies above the \HESS\ data. While a
better
fit could be obtained by reducing the external mass and/or by
increasing the CR diffusion coefficient in the CSM to reduce  the
contribution from escaping CRs, a good fit to the GeV-TeV emission
with only \pion\ isn't possible.
\label{fig:BbandB}}
\end{figure}

There are two important improvements over the broadband IC fit given
in Figure 4 of \citet{EPSR2010}. One is that the newer \Fermi\ data
for \SNRJmm\ now
clearly favor a IC model, whereas the preliminary
\Fermi\ data available for \citet{EPSR2010} were less clear.
The second improvement is in our match to the highest energy \HESS\
points. In the constant ISM model used in \citet{EPSR2010}, the IC
fit fell below the highest energy \HESS\ points. Now, with our \CC\
model, we are able to fit the highest energy points successfully with
only the CMB photon field. This
improvement comes about because the magnetic field at the FS is lower
in the \CC\ case and electrons can obtain a higher energy before
\synch\ losses dominate. Magnetic field amplification is still
important (we fit the data with $\Bamp=8.5$ for this case) but
starting with a lower ambient field is advantageous.

\subsection{External Molecular Cloud Interaction}\label{sec:MC}
The contribution from the escaping CRs to the TeV emission
will increase as the external target material increases, as would be
the case if the escaping CRs from the SNR diffused into a nearby
molecular cloud
\citep[e.g.,][]{AharA1996,PZ2005a,GA2007,
GAC2009,CAB2010,OhiraEtal2011}.
In  Figs.~\ref{fig:profB} and \ref{fig:BbandB}, we show an example
(Model B) where the mass external to the FS is $10^4\,\Msun$ with a
density of $\sim 100$\,\pcc. All other parameters are the same as for
Model A and again we note that our model is spherically symmetric so
all escaping CRs interact with the outer shell.
Now, the \pion\ emission from the escaping CRs at TeV energies is well
above the \pion\ from the trapped CR protons and is comparable at the
highest energies to the IC from the trapped CR electrons.

While a larger external mass would clearly result in \pion\
dominating the IC, it remains to be seen if a  satisfactory fit for
\SNRJmm\ can be found with just \pion.  The first problem concerns the
shape of the \pion\ emission from the escaping CR distribution. The
distributions shown in Figs.~\ref{fig:BbandA} and \ref{fig:BbandB} are
much too narrow to produce a good fit to both the \Fermi\ and \HESS\
fluxes and a substantial contribution from IC is required to produce a
good fit. We note, for example,
that \citet{ZP2008} and \citet{CAB2010} find similar
narrow escaping CR distributions, as does \citet{VEB2006} with Monte Carlo simulations that directly determine the escaping flux, but that \citet{GAC2009} and
\citet{OhiraEtal2011} assume
broader
distributions when averaged over the age of the 
remnant.\footnote{There are data from spacecraft observations of the Earth bow shock supporting the direct escape of a narrow distribution of accelerated particles 
\citep[e.g.,][]{ScholerEtal1980,MitchellEtal1983}.}
As was discussed in \citet{EB2011}, the shape of the escaping CR
particle distribution depends on the details of how the highest energy
particles, trapped and escaping, generate turbulence, 
particularly via long-wavelength effects.
The long-wavelength, wave-particle interactions that are important for these high-energy, escaping particles
involve highly
anisotropic distributions that are only beginning to be modeled  
\SCly\ \citep*[e.g.,][]{CAB2010,BOE2011,SchureBell2011}. It is certainly possible that broader
distributions of escaping CRs than we show may come from these
calculations or others,  and that a better fit to the GeV-TeV flux
without the IC component could be obtained. Nevertheless, we feel it
is unlikely that escaping CRs streaming away from a relatively young
SNR will have a distribution much broader then we show here.

Regardless of the details of the wave generation, however, the shapes
of the trapped CR ions and the escaping ones are related since the
turnover in the trapped ion spectrum comes about as CRs escape from
the FS. In the case of a high enough $B$-field, the shape of the
electron turnover is determined mainly by radiation losses.
With any attempt to broaden the spectral cutoff to allow a match to
the GeV-TeV emission, care must be taken so that the
X-ray \synch\ isn't broadened beyond acceptable limits. This is
particularly important for a \CC\ model where the effective magnetic
field is less than in a homogeneous model.

Another problem that must be addressed if \gamray\ production in
external material is important concerns the observed spatial
coincidence of \gamray\ and X-ray emission.
If \gamrays\ from pion production in external clouds were to dominate
the \gamray\ emission one would not expect the observed good agreement
between the X-ray and \gamray\ morphologies
\citep[i.e.,][]{AharonianJ1713_2006}.

\begin{figure}
\epsscale{1.1}
\plotone{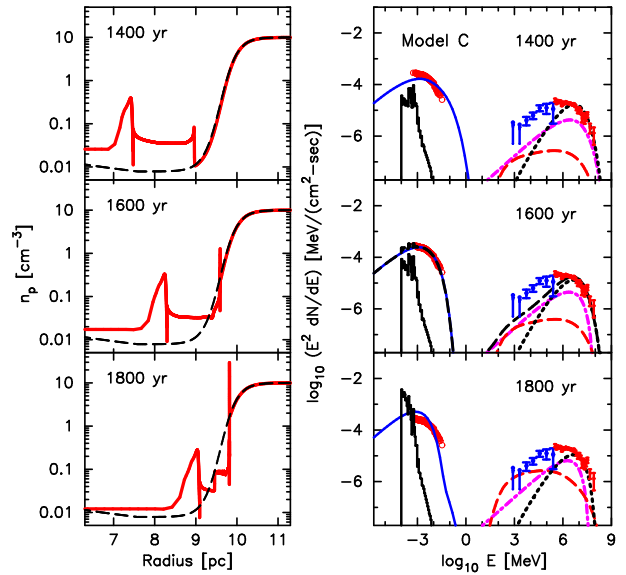}          % Fig 5
\caption{Impact of the FS on a $10^4\,\Msun$
spherical shell for
Model $C$. The left panels show the density profiles and the right
panels show the
high-energy emission at the ages indicated. As in
Figs.~\ref{fig:profA} and \ref{fig:profB}, the dashed black curves in
the left panels show the density profiles at the start of the
simulation and the solid red curves show the profiles at the ages
indicated. For the right-hand panels, the various emission processes
have the same  line characteristics as in Figs.~\ref
{fig:BbandA} and
\ref{fig:BbandB}, i.e., \syn\ (solid blue curve), IC (dot-dashed
purple curve),
\pion\ from trapped CRs (dashed red curve), \pion\ from escaping CRs (
dotted black curve), and thermal X-rays (solid black curve).
In the 1600\,yr right-hand panel we show the high-energy data along
with the summed emission (dashed black curve). While the
parameters
for the
1600\,yr case give a reasonable fit to the broadband, high-energy data
with \pion\ (from combined trapped and escaping CRs) dominating the
GeV-TeV emission, a better fit can only be obtained by increasing the
IC relative to the \pion. The thermal X-rays are calculated assuming
Coulomb
heating of electrons.
\label{fig:Time14_18}}
\end{figure}

\begin{figure}
\epsscale{0.85}
\plotone{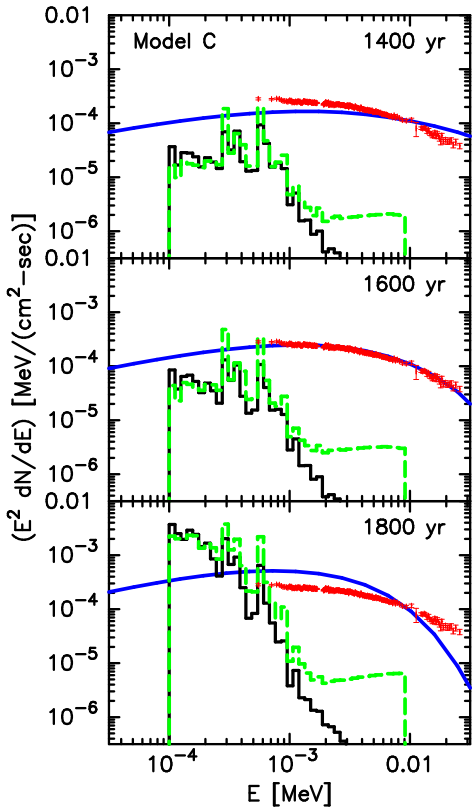}          % Fig 6
\caption{Time history of
X-ray
emission for the same Model $C$ as
shown in Fig.~\ref{fig:Time14_18}. In all
panels, the solid black curve shows thermal X-rays with Coulomb
electron temperature equilibration and the dashed green curve shows
thermal X-rays with instant equilibration. The solid blue
curve is the \syn\ continuum. We limit our thermal X-ray calculation
to energies between $10^{-4}$ and $10^{-2}$\,MeV, creating the  sharp
drop-offs beyond these energies.
 The Suzaku observations are shown in red.
\label{fig:Thermal}}
\end{figure}

\subsection{SNR Impacting External Material}\label{sec:MCimpact}
A possibility that has been discussed
 for \SNRJmm\ \citep[e.g.,][]{BV2006,BV2010} is that the FS is
 currently
 impacting a pre-SN shell or molecular cloud
at a radius of $\sim 10$\,pc. In this case, time scales considerably
less than $\tSNR$ become important
as the FS moves into the steep density ramp of the shell or cloud
edge. Since there is a time delay between the shock heating of the
plasma and the
non-equilibrium production of X-ray emission lines, the question
naturally arises: is it possible for GeV-TeV emission to be produced
before strong X-ray emission lines are generated?

In Fig.~\ref{fig:Time14_18}
 we show our model
C as a time sequence of the SNR density profile and the photon
emission as the FS runs into a density gradient. The parameters are
similar to model B except we have moved the inner edge of the dense
shell to $\sim 9.6$\,pc  so that the SNR radius is
about 10\,pc at 1600\,yr to be consistent with observations. We have
also lowered $\Kep$ to $10^{-3}$ and increased the wind speed to
30\,\kmps\ from 20\,\kmps\ and made other changes
(see Tables~\ref{tab:tableSNR}, \ref{tab:tableDSA} and
\ref{tab:tableCSM}) in the input
parameters in the attempt to obtain a good fit to the X-ray and
GeV-TeV observations at 1600\,yr
with \pion, from a combination of trapped and escaping CR protons,
dominating the GeV-TeV emission at $\tSNR=1600$\,yr.
As is clear from
Figs.~\ref{fig:BbandA} and \ref{fig:BbandB}, the mass concentrated in
the dense shell must be $\gg 100\,\Msun$ to make the \pion\
contribution from the escaping CRs comparable to IC and we use
$\Mshell=10^4\,\Msun$ for Fig.~\ref{fig:Time14_18}.

In Fig.~\ref{fig:Time14_18} the black dashed curves in the left-hand
plots are the
initial density profile and the red solid curves are the density
profiles at the ages shown. The right-hand plots show the broadband
emission at
the indicated ages.
The important properties of Fig.~\ref{fig:Time14_18} are:
\newlistroman

 \listromanDE in order to not overproduce the thermal X-rays,
 the density upstream  of the FS at $\sim 1600$\,yr must be
$\lsim 1$\,\pcc, forcing the FS to be ascending the density ramp and
not in the high-density plateau;

 \listromanDE the \pion\ emission from the escaping CRs varies weakly
 with $\tSNR$ since it depends largely on the constant amount of
 external mass in the dense shell;

 \listromanDE the \pion\ emission from the trapped CRs increases
 rapidly as the FS ascends the dense shell because it scales
 $\sim n_p^2$;

 \listromanDE as the FS moves into the ``\MCloud," the X-ray
\syn\ and TeV IC increase less rapidily than the \pion\ from the
trapped CRs because radiation losses increase due to the increasing
magnetic field in the cloud. This is particularly noticeable in the
\syn\ between 1400 and 1600\,yr;

 \listromanDE between 1600 and 1800\,yr, the relative intensity of
 thermal X-rays increases by more than a factor of ten relative to the
 nonthermal continuum emission;
and

\listromanDE the distance between the forward and reverse shocks
 drops substantially as the FS runs up the density gradient.

%% \newlistroman
%%  \listromanDE in order to not overproduce the thermal X-rays,
%%  the
%%  density upstream  of the FS at $\sim 1600$\,yr must be
%% $\lsim 1$\,\pcc, forcing it to be ascending the density ramp and
%% not in the high-density plateau;
%% \listromanDE the \gamray\ intensity from the trapped CR protons
%% (red
%%  dashed curves) increases more rapidily than that from the escaping
%%  CRs (black dotted curves)
%% as the FS first moves into the dense \MCloud;
%% \listromanDE from 1400 to 1700 years, the thermal X-ray emission
 %% barely increases in overall intensity even though the FS is
 %% encountering a region that has increased by nearly a factor of 100
 %% in
 %% density;
%% \listromanDE between 1700 and 1800 years, the thermal X-ray
%% emission
%%  jumps by nearly a factor of 10 in overall intensity and we
%% attribute
%%  this to the time delay needed for the non-equilibrium
%% thermal emission to be produced;
%% \listromanDE beyond 1800 years, the thermal X-ray emission
%% continues
%%  to rise above the X-ray \syn\ and the distribution of thermal
%%  emission lines tends to peak more sharply;

In Fig.~\ref{fig:Thermal} we show
the thermal X-rays along with the
\syn\ emission (solid blue curves) in the X-ray band for Model C.  The
solid black histograms are calculated assuming that electrons
immediately behind the shock are cold and then heated via Coulomb
collisions with the shock-heated protons. The dashed green histograms
are calculated assuming the electrons are heated instantly  with the
protons.
While the method of electron heating produces important differences in
the details of the emission lines, all major conclusions for broadband
models are independent of this heating \citep[see][for a
full discussion of our model for electron
heating]{EPSBG2007,EPSR2010}.
The calculation of the emission lines includes the non-equilibrium
effects as the plasma is shock heated and compressed and then expands
and cools behind the FS.
For this set of parameters, until $\sim 1600$ years, the emission
lines lie mainly at or below the
\syn\ continuum while at 1800\,yr, the line flux is more than 10 times
the \syn\ flux at $\sim 0.1$\,keV.
The variation in line intensity between $0.2$ and 1\,keV over the
400\,yr period
from 1400 to 1800\,yr is greater than an order of magnitude, while the
variation in \syn\ intensity in this energy range is less than a
factor of three.

\begin{figure}
\epsscale{0.995}
\plotone{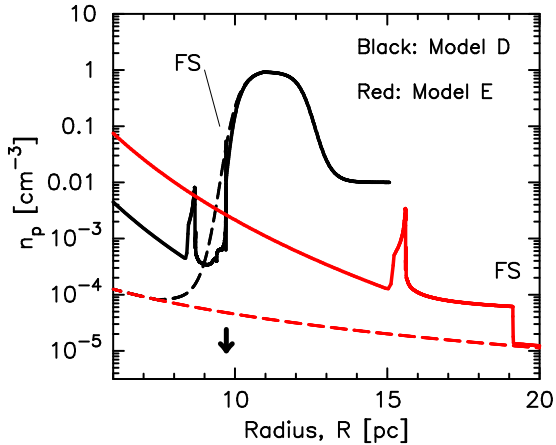}          % Fig 7
\caption{The red curves show the results at
$\tSNR=1630$\,yr for a fast pre-SN wind where any high-density shell
is at a radius beyond 20\,pc. The model producing the black curves has
exactly the same parameters only now the dense shell is placed at
$\Rshell \sim 9.5$\,pc. In this case, the FS impacts the shell well
before 1630\,yr and is shown 800\,yr after the explosion. The
arrow on the horizontal axis indicates the FS position at 800\,yr for
Model D. Since photon spectra are not calculated for Models D and E, only values important for the density profile are given in Tables~\ref{tab:tableDSA}, \ref{tab:tableCSM}, 
and \ref{tab:tableOUT}.
\label{fig:ShellDist}}
\end{figure}

Also evident in Fig.~\ref{fig:Thermal} is that the high-energy
\syn\ emission decreases with increasing age. We note that
\citet{PatnaudeCasA2011} have observed a decline in
the nonthermal emission from \CasA\ on time scales $\lsim 10$\,yr with
a 1.5\%--2\% yr$^{-1}$ decline in the 4.2--6.0 keV range. Such rapid
changes in \syn\ emission could result from a number of causes as the
FS enters a steep density gradient. These including the  slowing of
the FS, an increase in the ambient magnetic field strength, as is the
case in Fig.~\ref{fig:Time14_18}, and/or the damping of magnetic
turbulence in the weakly ionized dense material.

\subsection{High-Velocity Pre-SN Wind}
Another possibility is that \SNRJmm\ exploded in a fast wind with
$\Vwind \sim 1000$\,\kmps, and $\dMdt = 10^{-7}$\,\SunMyr,
typical of an
early B-type or late O-type star \citep[e.g.,][]{deJager88}.
If we accept the constraints that the remnant currently has an age of
$\sim 1600$\,yr, a radius of $\lsim 10$\,pc, and an
 explosion
energy of
$\EnSN \sim 10^{51}$\,erg, it's clear that the only possibility is
that the FS is now impacting dense material which stands $\sim 10$\,pc
from the explosion center.

To illustrate this, we show in Fig.~\ref{fig:ShellDist},
examples (Models D and E), where
$\Vwind=500$\,\kmps, $\dMdt = 10^{-6}$\,\SunMyr, and
$\Mej \sim 10\,\Msun$. The position of the FS at 1600\,yr will be well
beyond 10\,pc unless the FS is slowed by dense material. We note that
we have chosen extreme values for Fig.~\ref{fig:ShellDist}; a faster
$\Vwind$, a lower $\dMdt$, and/or a smaller $\Mej$, will result in a
FS beyond 20\,pc at 1600\,yr.
The slower speed we use also encompasses the case where the CSM contains
regions of both high and low speed pre-SN winds.

The black curves in Fig.~\ref{fig:ShellDist} show the case where the
shell edge is at $\Rshell \sim 9.5$\,pc, while the red curves show the
results for the same input parameters except that the dense shell is
beyond the radius the FS obtains at $\tSNR=1630$\,yr.
The same arguments used to exclude the FS penetrating the dense shell
in Fig.~\ref{fig:Time14_18} apply for a fast wind. The constraints of
age, distance, and radius are not consistent with the lack of X-ray
line emission if the FS is interacting with a dense shell.

\section{Discussion and Conclusions}
We have generalized our \CRnei\ model of an evolving SNR to include
the production of radiation in
a non-homogeneous CSM such as would exist with a pre-SN wind or for a
SN exploding near a \MCloud.
The model includes the production and escape of high-energy CR
protons, as well as CR electrons and protons that remain trapped
within the FS. As the escaping CR protons diffuse in the CSM, they
produce \gamrays\ by \pion\ and this production is added to the \syn,
\IC, and \pion\ emission from the trapped CRs. As in our previous work
\citep[e.g.,][]{EPSBG2007,PES2009,EPSR2010}, we simultaneously
calculate the thermal X-ray line emission with the broadband continuum
allowing  strong constraints to be placed on some of
the many parameters required to model the \SNRJmm.

Cosmic-ray protons that escape the SNR and interact with external
material produce  \gamrays\ without producing thermal X-rays.
If escaping CRs are not considered, homogenenous models clearly favor
IC from trapped CR electrons as the mechanism responsible for the
GeV-TeV emission in \SNRJmm. Including escaping CRs opens  the
possibility
that the constraint on ambient density imposed by the lack of X-ray
line emission in \SNRJmm\ can be satisfied and still have the GeV-TeV
emission produced predominately by CR protons in the external
material.

As in our previous work, we don't calculate the 
thermal X-ray emission
at the remnant reverse shock (RS), nor do we explicitly include the acceleration
of elements heavier than protons at the FS. We do include a 10\% by number contribution of helium in the target nuclei.  
%Heavy elements can contribute to shock modification and , although we %do include the
%effect of helium in the target nuclei for \pion. 
The calculations of \citet*{CBA2011} include heavy nuclei in the shock modification and show an increase in the \pion\ emission by
as much as a factor of 2.5 along with changes in the shape of the emitted
spectrum. However, the enhancement of \pion\ emission from heavy elements is not enough to modify our main conclusion that
leptons dominate the
GeV-TeV emission, even without considering the thermal emission from
the RS. Including the RS thermal emission would produce an even
greater enhancement of IC over \pion\ at GeV-TeV energies.

We have considered four cases. The first is when the external material
is a shell from a slow
pre-SN wind, in which case the external mass is $<100\,\Msun$. The
second is when a larger external shell of $\Mshell = 10^4\,\Msun$ lies
near but outside of the FS at $\tSNR=1630$\,yr. The third case
examines what happens when the FS is interacting with the dense
$10^4\,\Msun$ shell.
The fourth case considers a SNR in a fast pre-SN wind interacting with
a $100\,\Msun$ shell.

Our major conclusion is that it is not possible to
obtain a satisfactory broadband fit to \SNRJmm\ dominated by \pion\ if
the remnant  is isolated.
An excellent  fit can be obtained
(Fig.~\ref{fig:BbandA}) with IC
dominating the GeV-TeV emission, confirming that an isolated \SNRJmm,
whether from a core-collapse or thermonuclear SN, produces most of its
GeV-TeV emission by IC radiation.
It is possible, but unlikely, for \pion\ from escaping CRs to dominate
the GeV-TeV emission if  this remnant is near or interacting with
$\gsim 10^4\,\Msun$ of external material.

If the SNR is near a massive region $\sim 10^4\,\Msun$, then \pion\
from escaping CRs can, in principle, outshine the \IC\ from trapped CR
electrons at
TeV energies.
However, there are fundamental uncertainties in the properties of
escaping CRs having to do with  the shape of the escaping CR
distribution and their diffusion in the partially ionized CSM.
The  model we use produces a distribution that is too narrow to match
the combined \Fermi\ and \HESS\ observations with \pion\ alone and a
substantial contribution from IC is required for a satisfactory fit
see Fig.~\ref{fig:BbandB}.
The highest energy CRs, whether trapped or escaping, will have highly
anisotropic distributions and substantial work is needed to  better
understand wave generation in such  distributions before a firm
prediction for the shape can be made.

Our conclusion that the GeV-TeV emission from an isolated
\SNRJmm\ must
be dominated by IC is essentially independent of the CR diffusion
parameters in the CSM since escaping CRs contribute an insignificant
fraction of the emission for $\Mshell \le 100\,\Msun$. However, if the
escaping CRs are interacting with enough mass ($\gsim 10^4\,\Msun$),
the
diffusion properties will be important. In that case, the energy
dependence of $\DCSM$, as well as the normalization, will
influence
the shape and intensity of the \pion\ emission. As the CRs diffuse
through the molecular cloud, $\DCSM$ will be determined by
wave-particle interactions in a dense, partially ionized medium 
\citep[e.g.,][and references therein]{RevilleEtal2007}.
A detailed analysis of this propagation is beyond the scope of this paper. 

A new aspect of our work is the modeling of the remnant as the FS
impacts a density gradient as shown in Figs.~\ref{fig:Time14_18}
and \ref{fig:Thermal}. These figures show that the
emission, and particularly the relative thermal and non-thermal
fluxes, will vary substantially on short time scales $\sim 100$\,yr
depending on how long the shock has been interacting with the cloud.
We note that an important
approximation of our \CRnei\ model will modify this conclusion
somewhat. As the FS runs into a steep density gradient, the mass flux
crossing the shock will increase and, depending on the change in shock
parameters,  more mass will be injected into the DSA mechanism. In
reality, it will take
some time for this additional mass to be accelerated to TeV energies
but we assume the acceleration occurs essentially instantaneously.
The X-ray line emission is calculated consistently with time so the
relative  increase in thermal over TeV emission shown
in Fig.~\ref{fig:Time14_18} would actually be greater than what is
shown.

Significantly, we have found that the \CC\ scenario presented here
affords a
better fit to the broadband \SNRJmm\ observations than
the homogeneous Type Ia SN model presented in \citet{EPSR2010}. The
lower pre-SN wind magnetic field present in the \CC\ model, yields a
lower post-shock field, even with strong MFA, allowing electrons
to be accelerated to higher energies than in the homogeneous ISM.
These higher energy electrons produce IC emission against the CMB
consistent with the highest energy \HESS\ points.

The low magnetic field value we find is not necessarily inconsistent
with the rapid time variations observed for \SNRJmm.
\citet{Uchiyama_J1713_2007} observed $\sim 1$\,yr variations in X-ray
\syn\ emission and interpreted this as the  radiation loss time scale
for TeV electrons in $\sim 1$\,mG fields. However, other estimates
yield lower values \citep[see references in][]{EV2008}, and there is
an alternative explanation for rapid time variations that doesn't
require such large fields. \citet*{BykovDots2008} showed that a
steeply falling electron distribution in a turbulent magnetic field
can produce intermittent \syn\ emission consistent with the \citet{
Uchiyama_J1713_2007} observations. In the \citet{BykovDots2008} model,
root-mean-square fields of 10's of \muG\ are  adequate to explain
this time variation and are consistent
with our findings here.

It is important to note that while we feel the case is all but closed
for \SNRJmm, we make no claim that all SNRs will show
IC dominance at GeV-TeV energies.
The relative brightness of \pion\
versus IC depends largely on environmental parameters and \pion\ may
well dominate in some SNRs, particularly if they show strong X-ray
line emission. It is also possible that, in a particular remnant,
high-energy emission in some areas will be dominated by IC and in
others by \pion.

A final important point is that even though IC from \rel\ CR electrons
dominates \pion\ from \rel\ ions, all of our models, and all \NL\
models we are aware of, show the efficient
acceleration of protons. Far more energy is put into \rel\ protons
than into electrons by our DSA models. Our work does not in any way
call into
question the widely held believe that SNRs are the primary sources of
CRs, at least up to the so-called knee at $10^{15-16}$\,eV.
The evidence that some individual SNRs produce CR ions with
high efficiency, while indirect, is compelling
\citep[see, for example,][]
{RE92,HRD2000,WarrenEtal2005,HelderVinketal2009,MC2011}.

\acknowledgments D.C.E. acknowledges support from NASA
grants
ATP02-0042-0006, NNH04Zss001N-LTSA, 06-ATP06-21, and NNX11AE03G.
D.P. and P.S. acknowledges partial support from NASA
Contract NAS8-03060 and
Grant NNX09AT68G.  A.M.B. was
supported in part by the Laboratory of Astrophysics with Extreme
Energy Release at the Politechnical University, St. Petersburg, 
RBRF grant OFIm 11-02-12082, and by the RAS Presidium Program.
D.C.E. and P.S. are grateful to Aspen Center for Physics where part of
this work was done when the authors were
participating in a summer  program.

% bbbb  Note: must have files:  aa.bst  and  aa.cls
\bibliographystyle{aa} % A&A style
%\bibliography{c:/a_a_TOP/bibTeX/bib_DCE}
\bibliography{bib_DCE}

\begin{thebibliography}{62}
\expandafter\ifx\csname natexlab\endcsname\relax\def\natexlab#1{#1}\fi

\bibitem[{{Abdo} {et~al.}(2011){Abdo}, {Ackermann}, {Ajello}, {Allafort},
  {Baldini}, {Ballet}, {Barbiellini}, {Baring}, {Bastieri}, {Bellazzini},
  {Berenji}, {Blandford}, {Bloom}, {Bonamente}, {Borgland}, {Bouvier},
  {Brandt}, {Bregeon}, {Brigida}, {Bruel}, {Buehler}, {Buson}, {Caliandro},
  {Cameron}, {Caraveo}, {Casandjian}, {Cecchi}, {Chaty}, {Chekhtman}, {Cheung},
  {Chiang}, {Cillis}, {Ciprini}, {Claus}, {Cohen-Tanugi}, {Conrad}, {Corbel},
  {Cutini}, {de Angelis}, {de Palma}, {Dermer}, {Digel}, {Silva}, {Drell},
  {Drlica-Wagner}, {Dubois}, {Dumora}, {Favuzzi}, {Ferrara}, {Fortin},
  {Frailis}, {Fukazawa}, {Fukui}, {Funk}, {Fusco}, {Gargano}, {Gasparrini},
  {Gehrels}, {Germani}, {Giglietto}, {Giordano}, {Giroletti}, {Glanzman},
  {Godfrey}, {Grenier}, {Grondin}, {Guiriec}, {Hadasch}, {Hanabata}, {Harding},
  {Hayashida}, {Hayashi}, {Hays}, {Horan}, {Jackson}, {J{\'o}hannesson},
  {Johnson}, {Kamae}, {Katagiri}, {Kataoka}, {Kerr}, {Kn{\"o}dlseder}, {Kuss},
  {Lande}, {Latronico}, {Lee}, {Lemoine-Goumard}, {Longo}, {Loparco},
  {Lovellette}, {Lubrano}, {Madejski}, {Makeev}, {Mazziotta}, {McEnery},
  {Michelson}, {Mignani}, {Mitthumsiri}, {Mizuno}, {Moiseev}, {Monte},
  {Monzani}, {Morselli}, {Moskalenko}, {Murgia}, {Naumann-Godo}, {Nolan},
  {Norris}, {Nuss}, {Ohsugi}, {Okumura}, {Orlando}, {Ormes}, {Paneque},
  {Parent}, {Pelassa}, {Pesce-Rollins}, {Pierbattista}, {Piron}, {Pohl},
  {Porter}, {Rain{\`o}}, {Rando}, {Razzano}, {Reimer}, {Reposeur}, {Ritz},
  {Romani}, {Roth}, {Sadrozinski}, {Saz Parkinson}, {Sgr{\`o}}, {Smith},
  {Smith}, {Spandre}, {Spinelli}, {Strickman}, {Tajima}, {Takahashi},
  {Takahashi}, {Tanaka}, {Thayer}, {Thayer}, {Thompson}, {Tibaldo}, {Tibolla},
  {Torres}, {Tosti}, {Tramacere}, {Troja}, {Uchiyama}, {Vandenbroucke},
  {Vasileiou}, {Vianello}, {Vilchez}, {Vitale}, {Waite}, {Wang}, {Winer},
  {Wood}, {Yamamoto}, {Yamazaki}, {Yang}, \& {Ziegler}}]{AbdoJ1713_2011}
{Abdo}, A.~A., {Ackermann}, M., {Ajello}, M., {et~al.} 2011, \apj, 734, 28

\bibitem[{{Acero} {et~al.}(2009){Acero}, {Ballet}, {Decourchelle},
  {Lemoine-Goumard}, {Ortega}, {Giacani}, {Dubner}, \&
  {Cassam-Chena{\"i}}}]{Acero2009}
{Acero}, F., {Ballet}, J., {Decourchelle}, A., {et~al.} 2009, \aap, 505, 157

\bibitem[{{Adriani} {et~al.}(2011){Adriani}, {Barbarino}, {Bazilevskaya},
  {Bellotti}, {Boezio}, {Bogomolov}, {Bonechi}, {Bongi}, {Bonvicini},
  {Borisov}, {Bottai}, {Bruno}, {Cafagna}, {Campana}, {Carbone}, {Carlson},
  {Casolino}, {Castellini}, {Consiglio}, {De Pascale}, {De Santis}, {De
  Simone}, {Di Felice}, {Galper}, {Gillard}, {Grishantseva}, {Jerse},
  {Karelin}, {Koldashov}, {Krutkov}, {Kvashnin}, {Leonov}, {Malakhov},
  {Malvezzi}, {Marcelli}, {Mayorov}, {Menn}, {Mikhailov}, {Mocchiutti},
  {Monaco}, {Mori}, {Nikonov}, {Osteria}, {Palma}, {Papini}, {Pearce},
  {Picozza}, {Pizzolotto}, {Ricci}, {Ricciarini}, {Rossetto}, {Sarkar},
  {Simon}, {Sparvoli}, {Spillantini}, {Stozhkov}, {Vacchi}, {Vannuccini},
  {Vasilyev}, {Voronov}, {Yurkin}, {Wu}, {Zampa}, {Zampa}, \&
  {Zverev}}]{AdrianiEtal2011}
{Adriani}, O., {Barbarino}, G.~C., {Bazilevskaya}, G.~A., {et~al.} 2011,
  Science, 332, 69

\bibitem[{{Aharonian} {et~al.}(2006){Aharonian}, {Akhperjanian}, {Bazer-Bachi},
  {Beilicke}, {Benbow}, {Berge}, {Bernl{\"o}hr}, {Boisson}, {Bolz}, {Borrel},
  {Braun}, {Breitling}, {Brown}, {Chadwick}, {Chounet}, {Cornils},
  {Costamante}, {Degrange}, {Dickinson}, {Djannati-Ata{\"i}}, {O'C.~Drury},
  {Dubus}, {Emmanoulopoulos}, {Espigat}, {Feinstein}, {Fontaine}, {Fuchs},
  {Funk}, {Gallant}, {Giebels}, {Glicenstein}, {Goret}, {Hadjichristidis},
  {Hauser}, {Hauser}, {Heinzelmann}, {Henri}, {Hermann}, {Hinton}, {Hofmann},
  {Holleran}, {Horns}, {Jacholkowska}, {de Jager}, {Kh{\'e}lifi}, {Klages},
  {Komin}, {Konopelko}, {Latham}, {Le Gallou}, {Lemi{\`e}re},
  {Lemoine-Goumard}, {Lohse}, {Martin}, {Martineau-Huynh}, {Marcowith},
  {Masterson}, {McComb}, {de Naurois}, {Nedbal}, {Nolan}, {Noutsos}, {Orford},
  {Osborne}, {Ouchrif}, {Panter}, {Pelletier}, {Pita}, {P{\"u}hlhofer},
  {Punch}, {Raubenheimer}, {Raue}, {Rayner}, {Reimer}, {Reimer}, {Ripken},
  {Rob}, {Rolland}, {Rowell}, {Sahakian}, {Saug{\'e}}, {Schlenker},
  {Schlickeiser}, {Schuster}, {Schwanke}, {Siewert}, {Sol}, {Spangler},
  {Steenkamp}, {Stegmann}, {Superina}, {Tavernet}, {Terrier}, {Th{\'e}oret},
  {Tluczykont}, {van Eldik}, {Vasileiadis}, {Venter}, {Vincent}, {V{\"o}lk}, \&
  {Wagner}}]{AharonianJ1713_2006}
{Aharonian}, F., {Akhperjanian}, A.~G., {Bazer-Bachi}, A.~R., {et~al.} 2006,
  \aap, 449, 223

\bibitem[{{Aharonian} {et~al.}(2007){Aharonian}, {Akhperjanian}, {Bazer-Bachi},
  {Beilicke}, {Benbow}, {Berge}, {Bernl{\"o}hr}, {Boisson}, {Bolz}, {Borrel},
  {Braun}, {Brion}, {Brown}, {B{\"u}hler}, {B{\"u}sching}, {Carrigan},
  {Chadwick}, {Chounet}, {Coignet}, {Cornils}, {Costamante}, {Degrange},
  {Dickinson}, {Djannati-Ata{\"i}}, {O'C.~Drury}, {Dubus}, {Egberts},
  {Emmanoulopoulos}, {Espigat}, {Feinstein}, {Ferrero}, {Fiasson}, {Fontaine},
  {Funk}, {Funk}, {F{\"u}{\ss}ling}, {Gallant}, {Giebels}, {Glicenstein},
  {Gl{\"u}ck}, {Goret}, {Hadjichristidis}, {Hauser}, {Hauser}, {Heinzelmann},
  {Henri}, {Hermann}, {Hinton}, {Hoffmann}, {Hofmann}, {Holleran}, {Hoppe},
  {Horns}, {Jacholkowska}, {de Jager}, {Kendziorra}, {Kerschhaggl},
  {Kh{\'e}lifi}, {Komin}, {Konopelko}, {Kosack}, {Lamanna}, {Latham}, {Le
  Gallou}, {Lemi{\`e}re}, {Lemoine-Goumard}, {Lohse}, {Martin},
  {Martineau-Huynh}, {Marcowith}, {Masterson}, {Maurin}, {McComb}, {Moulin},
  {de Naurois}, {Nedbal}, {Nolan}, {Noutsos}, {Olive}, {Orford}, {Osborne},
  {Panter}, {Pelletier}, {Pita}, {P{\"u}hlhofer}, {Punch}, {Ranchon},
  {Raubenheimer}, {Raue}, {Rayner}, {Reimer}, {Reimer}, {Ripken}, {Rob},
  {Rolland}, {Rosier-Lees}, {Rowell}, {Sahakian}, {Santangelo}, {Saug{\'e}},
  {Schlenker}, {Schlickeiser}, {Schr{\"o}der}, {Schwanke}, {Schwarzburg},
  {Schwemmer}, {Shalchi}, {Sol}, {Spangler}, {Spanier}, {Steenkamp},
  {Stegmann}, {Superina}, {Tam}, {Tavernet}, {Terrier}, {Tluczykont}, {van
  Eldik}, {Vasileiadis}, {Venter}, {Vialle}, {Vincent}, {V{\"o}lk}, {Wagner},
  \& {Ward}}]{Aharonian2007}
{Aharonian}, F., {Akhperjanian}, A.~G., {Bazer-Bachi}, A.~R., {et~al.} 2007,
  \aap, 464, 235

\bibitem[{{Aharonian} {et~al.}(2011){Aharonian}, {Akhperjanian}, {Bazer-Bachi},
  {Beilicke}, {Benbow}, {Berge}, {Bernl{\"o}hr}, {Boisson}, {Bolz}, {Borrel},
  {Braun}, {Brion}, {Brown}, {B{\"u}hler}, {B{\"u}sching}, {Carrigan},
  {Chadwick}, {Chounet}, {Coignet}, {Cornils}, {Costamante}, {Degrange},
  {Dickinson}, {Djannati-Ata{\"i}}, {Drury}, {Dubus}, {Egberts},
  {Emmanoulopoulos}, {Espigat}, {Feinstein}, {Ferrero}, {Fiasson}, {Fontaine},
  {Funk}, {Funk}, {F{\"u}{\ss}ling}, {Gallant}, {Giebels}, {Glicenstein},
  {Gl{\"u}ck}, {Goret}, {Hadjichristidis}, {Hauser}, {Hauser}, {Heinzelmann},
  {Henri}, {Hermann}, {Hinton}, {Hoffmann}, {Hofmann}, {Holleran}, {Hoppe},
  {Horns}, {Jacholkowska}, {de Jager}, {Kendziorra}, {Kerschhaggl},
  {Kh{\'e}lifi}, {Komin}, {Konopelko}, {Kosack}, {Lamanna}, {Latham}, {Le
  Gallou}, {Lemi{\`e}re}, {Lemoine-Goumard}, {Lohse}, {Martin},
  {Martineau-Huynh}, {Marcowith}, {Masterson}, {Maurin}, {McComb}, {Moulin},
  {de Naurois}, {Nedbal}, {Nolan}, {Noutsos}, {Olive}, {Orford}, {Osborne},
  {Panter}, {Pelletier}, {Pita}, {P{\"u}hlhofer}, {Punch}, {Ranchon},
  {Raubenheimer}, {Raue}, {Rayner}, {Reimer}, {Reimer}, {Ripken}, {Rob},
  {Rolland}, {Rosier-Lees}, {Rowell}, {Sahakian}, {Santangelo}, {Saug{\'e}},
  {Schlenker}, {Schlickeiser}, {Schr{\"o}der}, {Schwanke}, {Schwarzburg},
  {Schwemmer}, {Shalchi}, {Sol}, {Spangler}, {Spanier}, {Steenkamp},
  {Stegmann}, {Superina}, {Tam}, {Tavernet}, {Terrier}, {Tluczykont}, {van
  Eldik}, {Vasileiadis}, {Venter}, {Vialle}, {Vincent}, {V{\"o}lk}, {Wagner},
  \& {Ward}}]{Aharonian_J1713_ERR2011}
{Aharonian}, F., {Akhperjanian}, A.~G., {Bazer-Bachi}, A.~R., {et~al.} 2011,
  \aap, 531, C1+

\bibitem[{{Aharonian} \& {Atoyan}(1996)}]{AharA1996}
{Aharonian}, F.~A. \& {Atoyan}, A.~M. 1996, \aap, 309, 917

\bibitem[{{Berezhko} {et~al.}(1996{\natexlab{a}}){Berezhko}, {Elshin}, \&
  {Ksenofontov}}]{BEK1996a}
{Berezhko}, E.~G., {Elshin}, V.~K., \& {Ksenofontov}, L.~T. 1996{\natexlab{a}},
  Journal of Experimental and Theoretical Physics, 82, 1

\bibitem[{{Berezhko} {et~al.}(1996{\natexlab{b}}){Berezhko}, {Elshin}, \&
  {Ksenofontov}}]{BEK1996b}
{Berezhko}, E.~G., {Elshin}, V.~K., \& {Ksenofontov}, L.~T. 1996{\natexlab{b}},
  Astronomy Reports, 40, 155

\bibitem[{{Berezhko} \& {V{\"o}lk}(2006)}]{BV2006}
{Berezhko}, E.~G. \& {V{\"o}lk}, H.~J. 2006, \aap, 451, 981

\bibitem[{{Berezhko} \& {V{\"o}lk}(2010)}]{BV2010}
{Berezhko}, E.~G. \& {V{\"o}lk}, H.~J. 2010, \aap, 511, A34

\bibitem[{{Blasi} \& {Amato}(2011)}]{BA2011a}
{Blasi}, P. \& {Amato}, E. 2011, ArXiv e-prints

\bibitem[{{Blasi} {et~al.}(2005){Blasi}, {Gabici}, \& {Vannoni}}]{BGV2005}
{Blasi}, P., {Gabici}, S., \& {Vannoni}, G. 2005, \mnras, 361, 907

\bibitem[{{Bykov} {et~al.}(2011){Bykov}, {Osipov}, \& {Ellison}}]{BOE2011}
{Bykov}, A.~M., {Osipov}, S.~M., \& {Ellison}, D.~C. 2011, \mnras, 410, 39

\bibitem[{{Bykov} {et~al.}(2008){Bykov}, {Uvarov}, \&
  {Ellison}}]{BykovDots2008}
{Bykov}, A.~M., {Uvarov}, Y.~A., \& {Ellison}, D.~C. 2008, \apjl, 689, L133

\bibitem[{{Caprioli} {et~al.}(2010){Caprioli}, {Amato}, \& {Blasi}}]{CAB2010}
{Caprioli}, D., {Amato}, E., \& {Blasi}, P. 2010, Astroparticle Physics, 33,
  307

\bibitem[{{Caprioli} {et~al.}(2011){Caprioli}, {Blasi}, \& {Amato}}]{CBA2011}
{Caprioli}, D., {Blasi}, P., \& {Amato}, E. 2011, Astroparticle Physics, 34,
  447

\bibitem[{{Caprioli} {et~al.}(2009){Caprioli}, {Blasi}, {Amato}, \&
  {Vietri}}]{CBAV2009}
{Caprioli}, D., {Blasi}, P., {Amato}, E., \& {Vietri}, M. 2009, \mnras, 395,
  895

\bibitem[{{Chevalier}(1999)}]{Chevalier99}
{Chevalier}, R.~A. 1999, \apj, 511, 798

\bibitem[{{Chevalier} \& {Luo}(1994)}]{CL94}
{Chevalier}, R.~A. \& {Luo}, D. 1994, \apj, 421, 225

\bibitem[{{de Jager} {et~al.}(1988){de Jager}, {Nieuwenhuijzen}, \& {van der
  Hucht}}]{deJager88}
{de Jager}, C., {Nieuwenhuijzen}, H., \& {van der Hucht}, K.~A. 1988, \aaps,
  72, 259

\bibitem[{{Drury}(2010)}]{Drury2010}
{Drury}, L.~O. 2010, ArXiv e-prints

\bibitem[{{Ellison} \& {Bykov}(2011)}]{EB2011}
{Ellison}, D.~C. \& {Bykov}, A.~M. 2011, ArXiv e-prints

\bibitem[{{Ellison} \& {Cassam-Chena{\"{\i}}}(2005)}]{EC2005}
{Ellison}, D.~C. \& {Cassam-Chena{\"{\i}}}, G. 2005, \apj, 632, 920

\bibitem[{Ellison {et~al.}(2004)Ellison, Decourchelle, \& Ballet}]{EDB2004}
Ellison, D.~C., Decourchelle, A., \& Ballet, J. 2004, A\&A, 413, 189

\bibitem[{{Ellison} \& {Eichler}(1985)}]{EE85}
{Ellison}, D.~C. \& {Eichler}, D. 1985, Physical Review Letters, 55, 2735

\bibitem[{{Ellison} {et~al.}(1990){Ellison}, {Moebius}, \& {Paschmann}}]{EMP90}
{Ellison}, D.~C., {Moebius}, E., \& {Paschmann}, G. 1990, \apj, 352, 376

\bibitem[{{Ellison} {et~al.}(2007){Ellison}, {Patnaude}, {Slane}, {Blasi}, \&
  {Gabici}}]{EPSBG2007}
{Ellison}, D.~C., {Patnaude}, D.~J., {Slane}, P., {Blasi}, P., \& {Gabici}, S.
  2007, \apj, 661, 879

\bibitem[{{Ellison} {et~al.}(2010){Ellison}, {Patnaude}, {Slane}, \&
  {Raymond}}]{EPSR2010}
{Ellison}, D.~C., {Patnaude}, D.~J., {Slane}, P., \& {Raymond}, J. 2010, \apj,
  712, 287

\bibitem[{{Ellison} \& {Vladimirov}(2008)}]{EV2008}
{Ellison}, D.~C. \& {Vladimirov}, A. 2008, \apjl, 673, L47

\bibitem[{{Fukui} {et~al.}(2003){Fukui}, {Moriguchi}, {Tamura}, {Yamamoto},
  {Tawara}, {Mizuno}, {Onishi}, {Mizuno}, {Uchiyama}, {Hiraga}, {Takahashi},
  {Yamashita}, \& {Ikeuchi}}]{Fukui2003}
{Fukui}, Y., {Moriguchi}, Y., {Tamura}, K., {et~al.} 2003, \pasj, 55, L61

\bibitem[{{Gabici} \& {Aharonian}(2007)}]{GA2007}
{Gabici}, S. \& {Aharonian}, F.~A. 2007, \apjl, 665, L131

\bibitem[{{Gabici} {et~al.}(2009){Gabici}, {Aharonian}, \&
  {Casanova}}]{GAC2009}
{Gabici}, S., {Aharonian}, F.~A., \& {Casanova}, S. 2009, \mnras, 396, 1629

\bibitem[{{Ginzburg} \& {Syrovatskii}(1964)}]{Ginzburg1964}
{Ginzburg}, V.~L. \& {Syrovatskii}, S.~I. 1964, {The Origin of Cosmic Rays},
  ed. {Ginzburg, V.~L.~\& Syrovatskii, S.~I.}

\bibitem[{{Helder} {et~al.}(2009){Helder}, {Vink}, {Bassa}, {Bamba}, {Bleeker},
  {Funk}, {Ghavamian}, {van der Heyden}, {Verbunt}, \&
  {Yamazaki}}]{HelderVinketal2009}
{Helder}, E.~A., {Vink}, J., {Bassa}, C.~G., {et~al.} 2009, Science, 325, 719

\bibitem[{{Hughes} {et~al.}(2000){Hughes}, {Rakowski}, \&
  {Decourchelle}}]{HRD2000}
{Hughes}, J.~P., {Rakowski}, C.~E., \& {Decourchelle}, A. 2000, \apjl, 543, L61

\bibitem[{{Inoue} {et~al.}(2009){Inoue}, {Yamazaki}, \& {Inutsuka}}]{Inoue2009}
{Inoue}, T., {Yamazaki}, R., \& {Inutsuka}, S. 2009, \apj, 695, 825

\bibitem[{{Katz} \& {Waxman}(2008)}]{KW2008}
{Katz}, B. \& {Waxman}, E. 2008, Journal of Cosmology and Astro-Particle
  Physics, 1, 18

\bibitem[{{Lee} {et~al.}(2008){Lee}, {Kamae}, \& {Ellison}}]{LKE2008}
{Lee}, S., {Kamae}, T., \& {Ellison}, D.~C. 2008, \apj, 686, 325

\bibitem[{{Meyer} {et~al.}(1997){Meyer}, {Drury}, \& {Ellison}}]{MDE97}
{Meyer}, J., {Drury}, L.~O., \& {Ellison}, D.~C. 1997, \apj, 487, 182

\bibitem[{{Mitchell} {et~al.}(1983){Mitchell}, {Roelof}, {Sanderson},
  {Reinhard}, \& {Wenzel}}]{MitchellEtal1983}
{Mitchell}, D.~G., {Roelof}, E.~C., {Sanderson}, T.~R., {Reinhard}, R., \&
  {Wenzel}, K. 1983, \jgr, 88, 5635

\bibitem[{{Moriguchi} {et~al.}(2005){Moriguchi}, {Tamura}, {Tawara}, {Sasago},
  {Yamaoka}, {Onishi}, \& {Fukui}}]{Moriguchi2005}
{Moriguchi}, Y., {Tamura}, K., {Tawara}, Y., {et~al.} 2005, \apj, 631, 947

\bibitem[{{Morlino} {et~al.}(2009){Morlino}, {Amato}, \& {Blasi}}]{MAB2009}
{Morlino}, G., {Amato}, E., \& {Blasi}, P. 2009, \mnras, 392, 240

\bibitem[{{Morlino} \& {Caprioli}(2011)}]{MC2011}
{Morlino}, G. \& {Caprioli}, D. 2011, ArXiv e-prints

\bibitem[{{Ohira} {et~al.}(2011){Ohira}, {Murase}, \&
  {Yamazaki}}]{OhiraEtal2011}
{Ohira}, Y., {Murase}, K., \& {Yamazaki}, R. 2011, \mnras, 410, 1577

\bibitem[{{Patnaude} {et~al.}(2009){Patnaude}, {Ellison}, \& {Slane}}]{PES2009}
{Patnaude}, D.~J., {Ellison}, D.~C., \& {Slane}, P. 2009, \apj, 696, 1956

\bibitem[{{Patnaude} {et~al.}(2011){Patnaude}, {Vink}, {Laming}, \&
  {Fesen}}]{PatnaudeCasA2011}
{Patnaude}, D.~J., {Vink}, J., {Laming}, J.~M., \& {Fesen}, R.~A. 2011, \apjl,
  729, L28+

\bibitem[{{Porter} {et~al.}(2006){Porter}, {Moskalenko}, \& {Strong}}]{PMS2006}
{Porter}, T.~A., {Moskalenko}, I.~V., \& {Strong}, A.~W. 2006, \apjl, 648, L29

\bibitem[{{Ptuskin} {et~al.}(2006){Ptuskin}, {Moskalenko}, {Jones}, {Strong},
  \& {Zirakashvili}}]{PMJSZ2006}
{Ptuskin}, V.~S., {Moskalenko}, I.~V., {Jones}, F.~C., {Strong}, A.~W., \&
  {Zirakashvili}, V.~N. 2006, \apj, 642, 902

\bibitem[{{Ptuskin} \& {Zirakashvili}(2005)}]{PZ2005a}
{Ptuskin}, V.~S. \& {Zirakashvili}, V.~N. 2005, \aap, 429, 755

\bibitem[{{Reville} {et~al.}(2007){Reville}, {Kirk}, {Duffy}, \&
  {O'Sullivan}}]{RevilleEtal2007}
{Reville}, B., {Kirk}, J.~G., {Duffy}, P., \& {O'Sullivan}, S. 2007, \aap, 475,
  435

\bibitem[{{Reynolds} \& {Ellison}(1992)}]{RE92}
{Reynolds}, S.~P. \& {Ellison}, D.~C. 1992, \apjl, 399, L75

\bibitem[{{Sano} {et~al.}(2010){Sano}, {Sato}, {Horachi}, {Moribe}, {Yamamoto},
  {Hayakawa}, {Torii}, {Kawamura}, {Okuda}, {Mizuno}, {Onishi}, {Maezawa},
  {Inoue}, {Inutsuka}, {Tanaka}, {Matsumoto}, {Mizuno}, {Ogawa}, {Stutzki},
  {Bertoldi}, {Anderl}, {Bronfman}, {Koo}, {Burton}, {Benz}, \&
  {Fukui}}]{Sano2010}
{Sano}, H., {Sato}, J., {Horachi}, H., {et~al.} 2010, \apj, 724, 59

\bibitem[{{Scholer} {et~al.}(1980){Scholer}, {Hovestadt}, {Klecker}, {Ipavich},
  \& {Gloeckler}}]{ScholerEtal1980}
{Scholer}, M., {Hovestadt}, D., {Klecker}, B., {Ipavich}, F.~M., \&
  {Gloeckler}, G. 1980, \grl, 7, 73

\bibitem[{{Schure} \& {Bell}(2011)}]{SchureBell2011}
{Schure}, K.~M. \& {Bell}, A.~R. 2011, ArXiv e-prints

\bibitem[{{Tanaka} {et~al.}(2008){Tanaka}, {Uchiyama}, {Aharonian},
  {Takahashi}, {Bamba}, {Hiraga}, {Kataoka}, {Kishishita}, {Kokubun}, {Mori},
  {Nakazawa}, {Petre}, {Tajima}, \& {Watanabe}}]{Tanaka2008}
{Tanaka}, T., {Uchiyama}, Y., {Aharonian}, F.~A., {et~al.} 2008, \apj, 685, 988

\bibitem[{{Uchiyama} {et~al.}(2007){Uchiyama}, {Aharonian}, {Tanaka},
  {Takahashi}, \& {Maeda}}]{Uchiyama_J1713_2007}
{Uchiyama}, Y., {Aharonian}, F.~A., {Tanaka}, T., {Takahashi}, T., \& {Maeda},
  Y. 2007, \nat, 449, 576

\bibitem[{{Vladimirov} {et~al.}(2006){Vladimirov}, {Ellison}, \&
  {Bykov}}]{VEB2006}
{Vladimirov}, A., {Ellison}, D.~C., \& {Bykov}, A. 2006, \apj, 652, 1246

\bibitem[{{Walder} {et~al.}(2011){Walder}, {Folini}, \& {Meynet}}]{walderea11}
{Walder}, R., {Folini}, D., \& {Meynet}, G. 2011, \ssr, 57

\bibitem[{{Warren} {et~al.}(2005){Warren}, {Hughes}, {Badenes}, {Ghavamian},
  {McKee}, {Moffett}, {Plucinsky}, {Rakowski}, {Reynoso}, \&
  {Slane}}]{WarrenEtal2005}
{Warren}, J.~S., {Hughes}, J.~P., {Badenes}, C., {et~al.} 2005, \apj, 634, 376

\bibitem[{{Zirakashvili} \& {Aharonian}(2010)}]{ZirA2010}
{Zirakashvili}, V.~N. \& {Aharonian}, F.~A. 2010, \apj, 708, 965

\bibitem[{{Zirakashvili} \& {Ptuskin}(2008)}]{ZP2008}
{Zirakashvili}, V.~N. \& {Ptuskin}, V.~S. 2008, \apj, 678, 939

\end{thebibliography}

\clearpage

% tttSNR
\begin{table}
\begin{center}
\caption{Input Parameters for SN and SNR.}
\label{tab:tableSNR}
\vskip6pt
\begin{tabular}{crrrrrrrrrrrr}
\tableline
\tableline
\\
Model\tablenotemark{a}\tablenotetext{1}{All models use $\dSNR=1$\,kpc,
$\EnSN=1\xx{51}$\,erg, and $n=7$.}
&$\tSNR$ &$\Mej$ &$\dMdt$ &$\Vwind$ &$\SigWind$ 
&$\Twind$\tablenotemark{b}\tablenotetext{2}{The unshocked temperature has very little influence on the solutions as long as it is $\lsim 10^6$\,K.}
\\
& [yr] & [$\Msun$] & [\SunMyr] & [\kmps] & & [K] \\
\\
\tableline
\\
A & 1630 & 3 & $1\xx{-5}$ & 20 & 0.03 & $10^4$ \\
B & 1630 & 3 & $1\xx{-5}$ & 20 & 0.03 & $10^4$ \\
C & 1800 & 3 & $6\xx{-6}$ & 30 & 0.03 & $10^4$ \\
D & 800 & 10 & $1\xx{-6}$ & 500 & $5\xx{-3}$ & $10^5$ \\
E & 1630 & 10 & $1\xx{-6}$ & 500 & $5\xx{-3}$ & $10^5$ \\
%
%%F & 1630 & 10 & $1\xx{-5}$ & 200 & $5\xx{-3}$ & $10^5$ \\
%
\tableline
\tableline
\end{tabular}
\end{center}
\end{table}
%above tttSNR

% tttDSA
\begin{table}
\begin{center}
\caption{Input Parameters for DSA and Line Emission.}
\label{tab:tableDSA}
\vskip6pt
\begin{tabular}{crrrrrrrrrrrr}
\tableline
\tableline
\\
Model\tablenotemark{a}\tablenotetext{1}{All models use  $\fsk=0.1$,
$\Ncut=30$, and a solar elemental composition.}
&$\EffDSA$ &$\Bamp$ &$\Kep$ &$\cutoff$ &$\tempEq$ &$\Norm$ \\
&[\%] & & & & & \\
\\
\tableline
\\
A & 25 & 8.5 & 0.01 & 0.75 & 1 & 0.75 \\
B & 25 & 8.5 & 0.01 & 0.75 & 1 & 0.65 \\
C & 50 & 10 & $10^{-3}$ & 0.75 & 0 or 1 & 0.9 \\
%
%%%D & 25 & 2.7 & 0.025 & 0.67 & 1 & --- \\
D & 25 & --- & --- & --- & --- & --- \\
%
%%%E & 25 & 2.7 & 0.025 & 0.67 & 1 & --- \\
E & 25 & --- & --- & --- & --- & --- \\
%
%%F & 25 & 2 & 0.01 & 0.67 & 1 & 1.0 \\
%
\tableline
\tableline
\end{tabular}
\end{center}
\end{table}
%above tttDSA

% tttCSM
\begin{table}
\begin{center}
\caption{Input Parameters for CSM and escaping CR diffusion.}
\label{tab:tableCSM}
\vskip6pt
\begin{tabular}{crrrrrrrrrrrr}
\tableline
\tableline
\\
Model\tablenotemark{a}\tablenotetext{1}{All models use
$\rgIndex=\nIndex=0.5$, and $\Nuni=0.01$\,\pcc.}
&$\DCSMz$ &$\Lz$ &$\Nshell$ &$\Mshell$ &$\Rshell$ &$\Bshell$ \\
& [cm$^2$s$^{-1}$] & [pc] & [\pcc] & [$\Msun$] & [pc] & [\muG] \\
\\
\tableline
\\
A & $1\xx{26}$ & $3.2\xx{-3}$ & 1 & 100 & 12 & 3 \\
B & $1\xx{26}$ & $3.2\xx{-3}$ & 100 & $10^4$ & 12 & 3 \\
C & $1\xx{27}$ & $0.032$ & 10 & $10^4$ & 9.6 & 1 \\
%
%%D & $1\xx{27}$ & $0.032$ & 1 & 100 & 9.5 & 1 \\
D & --- & --- & 1 & 100 & 9.5 & --- \\
%
%%E & $1\xx{27}$ & $0.032$ & 1 & 100 & $>20$ & 1 \\
E & --- & --- & 1 & 100 & $>20$ & --- \\
%
%%F & $1\xx{27}$ & $0.032$ & 1 & 100 & 9 & 1 \\
%
\tableline
\tableline
\end{tabular}
\end{center}
\end{table}
%above tttCSM

% tttOUT
\begin{table}
\begin{center}
\caption{Output Values at end of simulation.}
\label{tab:tableOUT}
\vskip6pt
\begin{tabular}{crrrrrrrrrrrr}
\tableline
\tableline
\\
Model &$\Rtot$
&$n_p$\tablenotemark{a}\tablenotetext{1}{Value just upstream of the FS
at the end of the simulation.}
&$B_0$\tablenotemark{a}
&$B_2$\tablenotemark{b}\tablenotetext{2}{Magnetic field just
downstream from the FS at the end of the simulation.}
&$\RFS$ &$\VelFS$ &$\Mswept$ &$\EnCR$ &$\EnEsc$ &$\DSAinj$ \\
& & [\pcc] & [\muG] & [\muG] & [pc] & [\kmps] & [$\Msun$] & [$\EnSN$]
& [$\EnSN$] & \\
\\
\tableline
\\
A & 4.6 & 0.013 & 0.22 & 10 & 8.9 & 4200 & 4.2 & 0.15 & 0.013
& 3.87 \\
B & 4.6 & 0.06 & 0.23 & 10 & 8.9 & 4200 & 4.2 & 0.15 & 0.013
& 3.87 \\
C & 5.6 & 2.0 & 0.7 & 75 & 9.8 & 850 & 4.3 & 0.17 & 0.03
& 4.0 \\
%
%%D & 4.6 & 0.01 & 0.4 & 10 & 9.7 & 2500 & 0.025 & $4\xx{-3}$
%%& $3\xx{-4}$ & 3.9 \\
D & 4.6 & 0.01 & --- & --- & 9.7 & 2500 & 0.025 & $4\xx{-3}$
& $3\xx{-4}$ & 3.9 \\
%
%%E & 4.6 & $1.5\xx{-5}$ & 0.07 & 1 & 19 & 9000 & 0.04 & $6\xx{-3}$
%%& $5\xx{-4}$ & 3.7 \\
E & 4.6 & $1.5\xx{-5}$ & --- & --- & 19 & 9000 & 0.04 & $6\xx{-3}$
& $5\xx{-4}$ & 3.7 \\
\tableline
\tableline
\end{tabular}
\end{center}
\end{table}
%above tttOUT

\end{document}